\documentstyle[aps,prl,floats,epsf]{revtex}
\newcommand{\no}{\nonumber}
\newcommand{\non}{\nonumber \\}
\newcommand{\ve}[1]{{\bf #1}}

\newcommand{\vrho}{\vec{\rho}}
\newcommand{\vomeg}{\vec{\omega}}

\newcommand{\be}{\begin{equation}}
\newcommand{\ee}{\end{equation}}
\newcommand{\bea}{\begin{eqnarray}}
\newcommand{\eea}{\end{eqnarray}}
\newcommand{\lp}{\left (}
\newcommand{\rp}{\right )}
\newcommand{\lb}{\left \{}
\newcommand{\rb}{\right \}}
\newcommand{\lbr}{\left [}
\newcommand{\rbr}{\right ]}
\newcommand{\ld}{\left .}
\newcommand{\rd}{\right .}
\newcommand{\rhok}{{\vec{\rho}}_{\ve{k}}}
\newcommand{\rhomk}{{\vec{\rho}}_{-\ve{k}}}

\newcommand{\mt}{m_{\tau}}

\newcommand{\mut}{\mu_{\tau}}
\newcommand{\mutb}{\mu_{\tau}+1}

\newcommand{\fr}{\frac{1}{2}}

\newcommand{\rhokf}{\vec{\rho}_{\ve{k_1}}\ldots\vec{\rho}_{\ve{k_{4}}}}

\newcommand{\delf}{\delta_{\ve{k_1}+\ldots+\ve{k_{4}},0}}

\def\EQN#1#2{{\begin{equation}\label{#2}#1\end{equation}}}
\def\QTR#1#2{{\csname#1\endcsname#2}}
\begin{document}
\title{Thermodynamic characteristics of the classical $n$-vector magnetic 
model in three dimensions} 
\author{ Z.E.Usatenko, M.P.Kozlovskii}

\address{\it 
 Institute for Condensed Matter Physics of the National Academy of 
Sciences of Ukraine 
\\ 290011 Lviv, 1 Svientsitskii Str.
\\ e-mail: pylyp@icmp.lviv.ua}
\vspace{-0.1cm}
\date{\today}
\maketitle
\begin{abstract}
The method of calculating the free energy and thermodynamic characteristics 
of the classical n-vector three-dimensional (3D) magnetic model at the 
microscopic level without any adjustable parameters is proposed. 
Mathematical description is perfomed using the collective variables (CV) 
method in the framework of the $\rho^4$ model approximation. The 
exponentially decreasing function of the distance between the particles 
situated at the N sites of a simple cubic lattice is used as the interaction 
potential. Explicit and rigorous analytical expressions for entropy,internal 
energy, specific heat near the phase transition point as functions of the 
temperature are obtained. The dependence of the amplitudes of the 
thermodynamic characteristics of the system for $T>T_c$ and $T<T_c$ on the 
microscopic parameters of the interaction potential are studied for the 
cases $n=1,2,3$ and $n\to\infty$. The obtained results provide the basis for 
accurate analysis of the critical behaviour in three dimensions including 
the nonuniversal characteristics of the system. 
\end{abstract}

\renewcommand{\theequation}{\arabic{section}.\arabic{equation}}
\section{Introduction.}
\setcounter{equation}{0}

\indent
Investigating the behaviour of the real three-dimensional systems near the 
phase transition (PT) point is one of the most important problems of 
condensed matter physics. In the present paper we propose a theoretical 
description scheme of the critical behaviour of the classical $n$-vector 
magnetic model in three dimensions on the microscopic level without any 
adjustable  parameters. The description is based on the original method of 
calculating the thermodynamic and structural characteristics of 3D model 
systems near the PT point, which is known as the collective variables (CV) 
method \cite{ref1},\cite{ref2}. Within the frame of an uniform scheme this 
method allowed to obtain universal characteristics of the system such as 
critical exponents and critical amplitude ratios for thermodynamic functions 
above and below the critical point as well as non-universal ones. Starting 
from the initial principles we calculate the critical temperature and obtain 
explicit analytical expressions for basic thermodynamic characteristics of 
the system as functions of the temperature and the $n$ component number of 
the model. For the first time the dependence of the thermodynamic 
characteristics of the system on the microscopic parameters of the initial 
interaction potential is investigated. The investigation of the 
non-universal properties of the system near the PT point require the use of 
non-Gaussian distributions of the spin density fluctuations. The suggested 
calculation of the partition function is based on the use of the simplest 
non-Gaussian measure density, i.t. the $\rho^4$ model \cite{ref3}.

We have devoted this paper to the widely studied 
classical $n$-vector magnetic model \cite{ref4} on three-dimensional 
simple cubic lattice, which is also known as the Heisenberg classical $O(n)$ 
spin model or, in field theoretic language, as the lattice $O(n)$ 
nonlinear $\sigma$ model. The investigation of the critical behaviour of 
the classical $n$-vector model and its partial cases was conducted in the 
frame of different methods such as high- and low- temperature series, the 
field theory, semi-microscopic scaling-field theory and Monte-Carlo 
calculations. The main attention in these works was given to the 
investigation of the universal properties of the system such as critical 
exponents \cite{ref5}-\cite{ref12} and some combinations of the critical 
amplitudes of thermodynamic functions \cite{ref13}-\cite{ref26}. Besides, 
the equation of state of Ising system was obtained to order $\epsilon^2$ by 
Avdejeva and Migdal \cite{ref27}, Bresin, Wallace and Wilson \cite{ref28}. 
Later these results were generalized for the case of the $n$-vestor model 
\cite{ref28}. 

Some important results were also obtained for calculating the thermodynamic 
functions near the critical point. One of the first works in this sphere is 
a paper by Wegner (the so-called Wegner's expansion)\cite{ref29}, 
suggesting the expression for the free energy with 
"irrelevant" operators in Wilson approach taken into account. The works by 
Fisher and Aharony \cite{ref30}, Nicoll and Albright \cite{ref31}, Nelson 
\cite{ref32} were also devoted to receiving of crossover scaling functions 
for temperatures $T>T_c$ in zero external field near four dimensions. In 
1974, Riedel and Wegner \cite{ref33} developed a numerical technique, 
termed the scaling-field method, for obtaining crossover scaling 
functions for the free energy and susceptibility. Crossover functions 
rather than power laws \cite{ref34}-\cite{ref36} have described the 
nonasymptotic region between criticality and the noncritical 
"background". In the frame of the massive field theory by Bagnuls and 
Bervillier \cite{ref10},\cite{ref37} the nonasymptotic critical behaviour 
for $d=3$ in the disordered phase case was analysed. They obtained 
explicit expressions for the correlation length $\xi$, the susceptibility 
$\chi$ and the specific heat $C$ as temperature functions in the 
disordered phase along the critical isohore for one-, two-, three - 
component models. The description of nonasymptotical (though still 
critical) behaviour was obtained as crossover between the Wilson-Fisher 
(near the critical temperature $T_c$) and mean-field (very far from 
$T_c$) behaviours with three adjustable parameters used.

At the present time the actual task of the critical phenomena physics is 
elaborating the methods giving quantity description of the critical 
behaviour of the system without using any adjustable parameters. This has 
been demonstrated, for example, in the works of Dohm and coworkers 
\cite{ref38}-\cite{ref40} in which calculation of the temperature dependence 
of the thermodynamic characteristics of the system was performed without 
using $\epsilon$ - expansion in the frame of the minimal subtraction scheme 
for the $n$- vector model in three dimensions on the basis of high-order 
perturbation theory and Borel resummation. In these papers the amplitude 
functions of the susceptibility, the correlation length, and the specific 
heat above and below $T_c$ up to two-loop order within the $\phi^4$ model 
for the cases $n=1, 2, 3$ were calculated. This approach exploits 
simultaneously the experimental information and simple relation between the 
specific heat above and below $T_c$ for defining the effective renormalized 
static coupling of the model in the terms of the measured specific heat. 
The effective renormalized static coupling determined in such a way allows 
to obtain expressions for other thermodynamic characteristics of the model 
above and below $T_c$ without using additional adjustable parameters. 
Besides, in the recent works by Butera and Comi \cite{ref41} high- and low- 
temperature expansions for the free energy, the susceptibility and the 
second correlation moment of the classical $n$-vector model on the simple 
cubic (sc) and the body centered cubic (bcc) lattices were extended to 
order $\beta^{21}$. This research only contains temperature dependence of 
the thermodynamic characteristics and does not give the possibility to 
describe the functional dependence of basic thermodynamic functions on the 
microscopic parameters of the interaction potential and characteristics of 
the crystal lattice. All this indicates that nonuniversal properties 
of the 3D system near the phase transition point have not been studied 
sufficiently yet. However, the precise role and significance of a lattice 
structure and interaction potential parameters for the approach to 
asymptotic critical behaviour still seems open to question. We hope that our 
explicit representations may provide useful benchmarks in studying this 
question. 

The approach to the investigation of the critical properties of the 
$n$-vector magnetic model, suggested in \cite{ref42} provides the necessary 
conditions for our complex approach to the study of the universal and 
non-universal phase transition characteristics. The approach suggested 
in this paper allows us to perform the analysis of the dependence of the 
thermodynamic characteristics of the $n$-vector $3D$ magnetic model in the 
vicinity of the phase transition point as functions of temperature and 
study their dependence on the microscopic parameters of the interaction 
potential and characteristics of the crystal lattice without any adjustable 
parameters being used. These results are interesting from the point of view 
of comparing the theoretical investigations and experimental data. 

In Sec.II of the present paper we perform the calculation of the partition 
function of the $n$-vector model using non-Gaussian measure density. The 
explicit analytical expressions for partial partition functions and general 
recursion relations (RR) between coefficients of the " effective Hamiltonian 
blocks" which arise in that case are obtained in the $\rho^4$-model 
approximation. We define the scope of application of these approximate 
solutions and show that RR as partial solution have a saddle-type fixed 
point for all $n$. We perform the calculation of the eigenvectors and 
eigenvalues of the RG transformation matrix and give the results of the 
investigation of the dependence of the PT temperature on the microscopic 
parameters of the interaction potential and the characteristics of the 
crystal lattice.

Sec.III is devoted to the calculation of the free energy of the $n$-vector 
magnetic model for temperatures above and below the phase transition  point. 
The main idea of such calculation lies in considering separately the 
contribution from the critical region (CR) where renormalization group 
symmetry takes place and the region of the long-wavelength fluctuations 
(LWF) of the spin moment density. It shown that in the case of the 
temperatures $T<T_c$ in the region of the LWF the fluctuations are described 
by non-Gaussian distribution with negative coefficient at square term.
The distributions of the spin moment density fluctuations after the 
selection of the ordering free energy is reduced to the Gaussian 
distribution. The dependence of the coefficients of the complete expression 
of the free energy on the microscopic parameters of the initial interaction 
potential and characteristics of the crystal lattice is investigated for the 
cases $n=1,2,3$. The contributions into the expressions for entropy and 
specific heat from the CR and LWF region are analyzed. It has been shown 
that considering the contribution of the LWF region satisfies the 
positiveness of the specific heat of the $n$ - vector model and system 
stability. 

In Sec.IV. the explicit expressions for thermodynamic functions of the model 
as functions of the temperature are obtained. The dependence of the critical 
amplitudes of the thermodynamic functions on the microscopic parameters of 
the interaction potential and the characteristics of the crystall lattice 
is investigated for the cases $n=1,2,3$. It has been shown that in the 
case $n=3$ (the Heisenberg model) the specific heat at the $T=T_c$ has the 
finite value. The dependence of the maximum of the specific heat on the 
microscopic parameters of the interaction potential is examined for the 
case $n=3$. The ratio of the critical amplitudes of the termodynamic 
functions at the temperatures $T>T_c$ and $T<T_c$ is calculated in order 
to compare the obtained results with the results of the other methods.

\renewcommand{\theequation}{\arabic{section}.\arabic{equation}}
\section{Calculation of the partition function and investigation of the 
recursion relations.}
\setcounter{equation}{0}

The critical behaviour of the different physical systems is characterized by 
their belonging to a specific class of universality which is defined by the 
dimensionality of the system $d$ and the symmetry of the order parameter 
$n$. The Stanley model \cite{ref4} is selected as the object of our 
investigation of the critical phenomena. This model describes the system of 
interacting $n$-component classical spins localized at the $N$ sites of the 
$d$-dimensional crystal lattice. The Stanley model is a generalization of 
the series of different models. In the case $n=0$ it is reduced to the task 
of a self-avoiding walk and is used for describing the polymerization 
phenomena. The cases $n=1,2,3$ correspond to the Ising model, the 
$XY$ - model and the Heisenberg model respectively. The boundary case  
$n\rightarrow \infty$ is equivalent to the Berlin-Kac spherical model 
\cite{ref43} for which the exact result is known. The Stanley model is 
described by the Hamiltonian 

\EQN{\QTR{bf}{\hat H}=-\frac 12\sum_{\QTR{bf}{R},\QTR{bf}{R}^{^{\prime
}}}J\left( \left| \QTR{bf}{R-R}^{^{\prime }}\right| \right) \QTR{bf}{\hat
S}_{\QTR{bf}{R}}\QTR{bf}{\hat S}_{\QTR{bf}{R}},}{f1}

where $\QTR{bf}{\hat S}_{\QTR{bf}{R}}=\left( \hat S_{\QTR{bf}{R}}^{\left(
 1\right) },...,\hat S_{\QTR{bf}{R}}^{\left( n\right) }\right) $
is the $n$ - component classical spin with the length $m$ $\left( 
\sum\limits_{\alpha=1}^n\left| \QTR{bf}{\hat 
S}_{\QTR{bf}{R}}^{(\alpha)}\right|^2=m^2\right) $. 
Spins are localized at the $N$ cites of the $d$ dimensional simple cubic 
lattice with coordinates $\vec R$. The interaction has an exchange character 
and can be described by the exponentially decreasing function of distance 
beetwen the particles.

\EQN{J\left( \left| \QTR{bf}{R-R}^{^{\prime }}\right| \right) =A_0\exp
\left(- \frac{\left| \QTR{bf}{R-R}^{^{\prime }}\right| }b\right) ,}{f2}

where $A_0, b$ are the constant. In the CV 
$\vrho_{\QTR{bf}{k}}=\left(\rho _{\QTR{bf}{k}}^{\left( 
     1\right) },...,\rho_{\QTR{bf}{k}}^{\left( n\right) }\right) $
representation the partition function of the model (\ref{f1}) is written as 
\cite{ref1},\cite{ref42}

 \EQN{Z=\int \exp \left[ \frac 12\sum_{\bf{k}}\beta \Phi (k) 
 \vrho_{\bf{k}}\vrho_{-\bf{k}}\right] J [\rho]  
 (d\vrho_{\bf{k}}) ^N\quad,}{f3}
where $\Phi (k)$ is the  Fourier-transform of the interaction potential 
(\ref{f2}), the element of the phase space is 
\footnote 
{The prime means that the product over $\bf{k}$ is maintained in the upper 
half-cube of the Brillouin zone and $k\ne 0$. The wave vector $\bf{k}$ 
assumes all the values inside the first Brillouin zone 
$$ k_{i}=2\pi n_{i}/N^{"}c, \quad i=1,..,d,$$
where $n_{i}$ are integers, 
$$\frac{-N^{"}}{2}\leq n_{i}<\frac{N^{"}}{2},\quad (N^{"})^{d}=N,$$
and $N$ is the number of the particles in the periodicity volume 
$V=Nc^{d}$, $c$ is a constant of the simple cubic lattice.} 
 
 $$(d\vrho_{\bf{k}}) ^N=\prod\limits_{a=1}^nd\rho
 _0^a\prod\limits_{\bf{k}}^{\prime }d\rho
 _{\QTR{bf}{k}}^{a,c}d\rho_{\QTR{bf}{k}}^{a,s},
 $$
and $J[\rho]$ is the transition Jacobian from the spin variables to the CV. 
The expression for $J[\rho]$ is given in Appendix 1. In the coordinate CV 
representation
\be
\vrho(\bf{R})=\frac{1}{\sqrt {N}}\sum_{\bf{k}} \vrho_{\bf{k}}\exp (-i 
\bf{kR}) \label{ddf4}
\ee
the expression for $J[\rho]$ is factored
\be
J[\rho]=\exp(u_{0}^{\prime}N^{\prime})J^{\prime}[0]\prod\limits_{\bf{R}}J[\rho(\bf{R})] 
\label{ddf5}\ee
where 
\be
J[\rho({\bf{R}})]=
\sum_{l\geq 1}\frac {a_{2l}}{(2l)!}|\vrho({\bf{R}})|^{2l}. 
\label{ddf6}
\ee
The expressions for $a_{2l}$ are shown in Appendix 2. The partition 
function (\ref{f3}) contains the contributions of two different types. The 
first is energy contributions 
\be
\frac{1}{2}\sum_{k\le B}\beta\Phi(k)\vrho_{\bf{k}}\vrho_{-{\bf{k}}}.
\label{ddf7}
\ee
They are diagonal in $\rho_{\bf{k}}$ representation and connected with the 
interaction potential. The second is entropy contributions 
\be
\sum_{l=1}^N \ln J(\rho({\bf{R}})). \label{ddf8}
\ee
They are diagonal in $\rho({\bf{R}})$ coordinate CV representation.
There are two possible approaches to calculating (\ref{f3}). The first one 
is using $\rho_{\bf{k}}$ variables in the calculation of the partition 
function of the system. In such a case the energy contributions are diagonal 
and the nondiagonality of the entropy contributions leads to the approximate 
method of calculation. One of such approximations is writing (\ref{f3}) in 
the following form
\bea
&&Z=\int \exp (-\frac{1}{2} \sum_{k\geq B}(a_{2}-\beta\Phi(k))\vrho_{\bf{k}}
\vrho_{-{\bf{k}}})\non
&&\times[1+\eta+\frac{1}{2}\eta^2+...](d\vrho_{\bf{k}})^N, 
\label{ddf9}
\eea
where  
\be
\eta=\sum_{l\geq 2} \frac{a_{2l}}{(2l)!} N^{1-l}
\sum_{k_{i}}\vrho_{{\bf{k}}_{1}},...\vrho_{{\bf{k}}_{2l}} 
\delta_{{\bf{k}}_{1}+...+{\bf{k}}_{2m}}.\label{ddf10}
\ee
As rule, such a way of calculation assumes that for the value $\eta$ 
it is sufficient to restrict the consideration to one term $(l=2)$ and in 
addition it is presupposed that \cite{ref3}
\be
(a_{2}-\beta\Phi(k))^{l} \gg a_{2l}. \label{ddf11}
\ee
The second way of the calculation of the partition function (\ref{f3}) 
assumes the use of the $\rho({\bf{R}})$ CV representation. In that case the 
entropy contributions are diagonal (\ref{ddf5}),(\ref{ddf6}) and 
approximation relates to the energy contribution (\ref{ddf7}), as it is 
nondiagonal in the $\rho({\bf{R}})$ representation. The essence of such 
fitting leads to a certain approximation of $\Phi(k)$ (see.\cite{ref1}) 
which allows to diagonalize the expressions in the exponent under the 
integrand for the partition function 
\bea 
&&Z=\prod \limits_{\bf{R}} \int \exp (-\frac{1}{2} (a_{2}-\beta \Phi_{apr})
\vrho({\bf{R}})^2 - \non
&&\sum_{l\ge 2} \frac{a_{2l}}{(2l)!} 
\vrho({\bf{R}})^{2l})d\vrho({\bf{R}}).\label{ddf12}
\eea
We use such approximatiuon for $\Phi_{apr}$ in which $\Phi_{apr}$ is a 
constant for every interval $k\in (B_{l},B_{l-1})$ and equals the respective 
mean value $\Phi(k)$. The first way corresponds to considering the moments 
of a certain Gaussian distribution. It allows us to calculate only certain 
classes of the graphs and does not solve the problem of the description of 
the critical behaviour as such. Besides, that condition (\ref{ddf11}) is too 
strong in the vicinity of the phase transition point. Nevertheless, such 
approach has its advantages at the expense of considering the wave vector  
dependence of $\Phi(k)$. The second way makes it impossible for us to 
consider the dependence of $\Phi(k)$ on the wave vector $k$ on the boundary 
of the Brillouin zone.\footnote
{The method considering the dependence on the wave vector ${\bf{k}}$ in 
the frame of the non-Gaussian measure density was proposed in the work
\cite{ref1}. It should be noticed that the $\eta$ correlation 
function critical exponent in the case of Ising model was calculated in 
these works.} 
Nevetheless, such a way is not restricted to Gaussian moments 
and is based on the use of the non-Gaussian measure density . In this case 
we do not need to perform the summation of various types of infinite series 
of perturbation theory, the certain terms of which tend to infinity with 
$T\to T_c$. Besides,the condition (\ref{ddf11}) becomes optional. In our 
opinion this condition is the basic barrier in the description of the 
critical behaviour of the three dimensional systems. The calculation of the 
nonuniversal characteristics of the phase transition particulaly the PT 
temperature $T_c$ is connected with the choice of the interaction potential. 
The Fourier - transform of the interaction potential (\ref{f2}) takes the 
form
 \EQN{\Phi(k) =\frac{\Phi(0) }{(1+b^2 k^{2}) ^2},}{f8}

where $\Phi \left( 0\right) =8\pi A_0\left( \frac bc\right) ^3$. The value 
of $\Phi(k)$ for the wave vectors similar to the boundary of the Brillouin 
half zone $(B=\frac \pi c)$ is much small than $\Phi(0)$. In this region of 
the wave vectors a weak dependence of $\Phi (k) $ on the wave vector is 
observed. This allows to accept the following approximation for $\Phi (k) $ 
 \EQN{\Phi (k) =\left\{\begin{array}{lcr}
 &&\Phi \left( 0\right) \left( 1-2b^2 k^{2}\right),  k<B^{^{\prime}}\\
 &&\bar \Phi =const , B^{^{\prime }}\leq k<B.
 \end{array}\right. }{f9}
The coordinate $B^{^{\prime }}$ is obtained from the condition of the 
applicability of the parabolic approximation for $\Phi(k) $ and equals 
 \EQN{B^{^{\prime }}=\left( b\sqrt{2}\right) ^{-1}.}{f10}

Among the set of the CV $\vrho_{\QTR{bf}{k}}$ there are those connected with 
order parameter. In the case of a model with the exponentially decreasing 
interaction potential (\ref{f2}) it is the $\vrho_0$ variable. The 
investigation of the critical behaviour of this model is largely determined 
by considering the contribution from the $\vrho_0$ variable in calculating 
the free energy. The mean value $\vrho_0$ describes the behaviour of the 
order parameter. Nevetheless, as seen from the (\ref{f3}),(\ref{f4}), all 
the CV $\vrho_{\bf{k}}$ are interconnected, and the contribution of the 
variable $\vrho_0$ alone cannot be separated in the partition function 
(\ref{f3}). The given task can be accomplished in case we use the method 
suggested in \cite{ref1}. Its essence lies in sequential integrating of the 
$\vrho_{\bf{k}}$ variables with $k\neq 0$ and the investigating of the 
functional from the $\vrho_0$ variable. In the case of the $n$-vector model 
such calculations are made for the first time. The functional representation 
of the partition function of the $n$ - vector magnetic model considering the 
interaction $\Phi(k) $ at $k$ close to the boundary of the Brillouin zone 
(\ref{f9}) has the form \cite{ref44}
\bea
&&Z=J^{'}[0] \exp(u_0^{\prime}N^{\prime}) \int \exp \left [-\frac 
{1}{2}\sum_{k<B^{\prime}}d(k)\vrho_{\bf{k}}\vrho_{\bf{-k}} - \right . \non
&&\left . -\frac{a_4}{4!N^{\prime}}\sum\limits_{{\bf{k}}_1,...,{\bf{k}}_4, 
k_i< B^{'}}\vrho_{{\bf{k}}_1}...\vrho_{{\bf{k}}_4}
\delta_{{\bf{k}}_1+...+{\bf{k}}_4}\right] 
(d\vrho_{\bf{k}})^{N^{\prime}}.\label{f25}
\eea
Part of the interaction potential is contained in the coefficient 
$$d (k) = a_2-\beta \Phi (k).$$ 
Besides, the coefficients $a_{2l}$ have as a small correction 
$\left(\bar \Phi <\Phi \left( 0\right) \right) $ that part of the 
Fourier-transform of the potential $\Phi (k) $ which corresponds to the 
values of $k$ in the vicinity of the Brillouin half zone (see Appendix 2).
The basic idea of the calculating (\ref{f25}) lies in a sequential exclusion 
from the consideration of the short-wavelength variables $\vrho_{\bf{k}}$
which decribe the behaviour of the effective spin blocks whose size is small 
in comparison with the system correlation length as the function of the 
temperature. After each of such step-by-step exclusion the sizes of 
effective spin blocks are increased by the factor of $s$ ($s\geq 1$). 
The set of the CV $\vrho_{\bf{k}}$ is divided into subsets. Each of these 
subsets contains the variables $\vrho_{\bf{k}}$ with certain values of the 
wave vectors $\bf{k}$. For the $l$ subset we have $ k\in (B_{l+1}B_l) $, 
where $B_{l+1}=\frac{B_l}s$, and $B_0=B^{^{\prime }}$. In each of the layers 
of the CV phase space the value $\Phi(k) $ is replaced by the corresponding 
mean value \cite{ref1}. After the layer integration the partition function 
(\ref{f25}) can be represented as a product of the partial partition 
functions $Q_{l}$ of separate layers 
 \be
 Z=C_l^{^{\prime }}Q_0Q_1...Q_l\int \left( d\vrho
 _{\QTR{bf}{k}}\right) ^{N_l+1}\omega _{l+1}\left( \vrho\right)
 ,\label{f43}
\ee
 where
  \bea
  \omega _{l+1}\left( \vrho\right) =\exp \left[
 -\frac 12\sum_{k< B_{l+1}}d^{\left(n, l+1\right)
 }(k) \vrho_{\QTR{bf}{k}}\vrho_{-\QTR{bf}{k}}-\right. \non
 \left. -\frac{a_4^{\left(n, l+1\right)
 }}{4!N_{l+1}}\sum_{{\bf{k}_1,...,\bf{k}_4},k_{i}<
 B_{l+1}}\vrho_{\QTR{bf}{k}_1}...\vrho
 _{\QTR{bf}{k}_4}\delta_{\QTR{bf}{k}_1+...+\QTR{bf}{k}_4}\right] ,\label{f44}
\eea
 \begin{eqnarray}
 && C_l^{^{\prime }}=\sqrt{2}^{\left( N_l-N_{l-1}\right) n},\ Q_0=\left[
 Q\left( u\right) Q\left( d_0\right) \right] ^{N^{^{\prime
 }}}, \nonumber \\
 && \ Q_l=\left[ Q\left( d_l\right) Q\left( P_{l-1}\right) \right] ^{N_l}.
 \label{f45}
 \end{eqnarray}
The value $ln\omega _{l+1}\left( \vrho\right) $ corresponds to the $l$ - 
block Hamiltonian which depends on the $N_l$ variables $\vrho_{\QTR{bf}{k}}$ 
connected with the fluctuation of the spin moment density in the blocks. The 
values which are part of the expressions for the partial partition functions 
are written in the form 

 $$Q\left( d_l\right) =\left( 2\pi \right) ^{\frac n2}\left( \frac
 3{a_4^{\left(n,l\right) }}\right) ^{\frac n4}U\left(
 \frac{n-1}2,x_l \right) \exp \left( \frac{x_l^{2}}{4}\right)
 ,$$
 \begin{eqnarray}
 && Q\left( P_{l-1}\right) =\left( 2\pi \right) ^{-\frac
 n2}\left(\frac{n+2}3s^d\frac{a_4^{\left(n,l-1\right) }}{\varphi_n
 x_{l-1}}\right) ^{\frac n4}\times \nonumber \\
 && \times U\left( \frac{n-1}2,y_{l-1}\right) \exp \left(
 \frac{y_{l-1}^2}4\right) , \label{f46}
 \end{eqnarray}
 \begin{eqnarray}
 && Q^{N^{^{\prime }}}\left(u\right) =J^{^{\prime }}\left[ 0\right] \exp
 \left( u_0^{^{\prime }}N^{^{\prime }}\right) .\label{f47}
 \end{eqnarray}
For coefficients $a_2^{\left(n,l\right) }$ and $a_4^{\left(n,l\right) }$ the 
recursion relations (RR) are valid
 \begin{eqnarray}
  && a_2^{\left(n,l+1\right) }=a_2^{\left(n,l\right) }+d^n\left(
  B_{l+1},B_l\right) M_n\left( x_l \right) ,\nonumber \\
  && a_4^{\left(n,l+1\right) }=a_4^{\left(n,l\right) }s^{-d}E_n\left(
  x_l \right) , \label{f48}
  \end{eqnarray}

where $d^n\left( B_{l+1},B_l\right) $ corresponds to the mean value in the 
$l$ layer. Specific functions introduced here have the following form
\be
 E_n\left( x_l\right) =s^{2d}\frac{\varphi _n\left(
 y_l\right) }{\varphi _n\left( x_l\right) },\quad  
 M_n\left( x_l\right) = N_n\left( x_l \right) -1,\quad
 N_n\left( x_l\right) = \frac{y_l U_n\left(
 y_l\right) }{x_l U_n\left( x_l\right) },\label{f49}
 \ee
where the functions $\varphi_n(t)$, $U_n(t)$ are combinations of the 
parabolic cylinder functions $U(a,t)$
$$
\varphi_n(t)=(n+2) U^{2}_n(t) + 2t U_n (t) - 2,\quad U_n 
(t)=\frac{U(\frac{n+1}{2},t)}{U(\frac{n-1}{2},t)}.
$$
The variables 
 \begin{eqnarray}
 && x_l=\sqrt{\frac 3{a_4^{\left(n,l\right) }}}d^n\left(
 B_{l+1},B_l\right) ,\nonumber \\
 && \ y_l=s^{\frac d2}U_n\left(x_l\right)
 \left( \frac{n+2}{\varphi _n\left( x_l\right) }\right)
 ^{\frac 12}. \label{f50}
 \end{eqnarray}
serve as arguments, where $N_l=N^{^{\prime }}s^{-dl}$, $d$ is the dimension 
of space, $l$ is the layer of integration. The obtained expression 
(\ref{f43}) for the partition function enables us to calculate the free 
energy of the system
 \EQN{F=-kT\sum_{l\geq 1}\ln Q_l\quad .}{f51}
 
when the explicit analytical expression for the partial partition function 
$Q_l$ (\ref{f45}), (\ref{f46}) are familiar to us. But, the calculation of 
the (\ref{f51}) can only be done if the explicit solutions of the RR 
(\ref{f48}) are obtained as functions of the phase layer number $l$. Similar 
investigation of the RR for the $n$ - vector model was carried out in series 
of works \cite{ref42},\cite{ref45},\cite{ref46},\cite{ref47}. The main 
attention in these papers was given to calculating the correlation 
function critical exponent and other universal characteristics of the 
model for the marginal cases 
 $s\rightarrow 1$ \cite{ref42},\cite{ref45},\cite{ref47} and $s\gg 1$ 
 \cite{ref46}. In this work we briefly present the results of the 
 investigation of the RR difference form (with $s>1$) \cite{ref44} which 
 allow us to perform the calculation of the non-universal system 
 characteristics. We introduce the following definitions 
\be
r_1^{(n)}=s^{2l}d^n(0),\quad u_1^{(n)}=s^{4l}a_4^{(n,l)}, \label{g1}
\ee
with 
\be
d^n(B_{l+1},B_l)=\frac{r_l^{(n)}+q}{s^{2l}}, \label{g2}
\ee
where 
\be
q=\beta\Phi(0)\bar q,~~~~\bar q=\frac{d}{(d+2)}\frac{(1-s^{-(d+2)})}
  {(1-s^{-d})}, \label{g3}
\ee
$\bar q$ is the geometric mean value of $k^2$ in the interval 
$(\frac{1}{s},1)$. As a result, the obtained RR (\ref{f48}) have the 
following representation 
\bea
&& r_{l+1}^{(n)}=s^2\lbr-q+(r_l^{(n)}+q)N_n(x_l)\rbr,\non
&& u_{l+1}^{(n)}=s^{4-d}u_l^{(n)}E_n(x_l).\label{g4}
\eea
The equations (\ref{g4}) as a partial solution have a fixed point 
\be
r_n^*=-f_n\beta\Phi(0), u_n^*=\varphi_n(\beta\Phi(0))^2, \label{g5}
\ee
where 
\bea
&& f_n=\frac{s^2(N_n(x^*)-1)}{s^2N_n(x^*)-1}\bar q,\non
&& \varphi_n=\frac{3}{(x^*)^2}\bar q^2\lb \frac{1-s^{-2}}{N_n
  (x^*)-s^{-2}}\rb^2. \label{g6}
\eea
The constants $f_n$ and $\varphi_n$ depend on the component number $n$ of 
the model and the universal value $x^*$ which is a solution of the nonlinear 
equation 
\be
s^{4+d}\varphi_n(y(x^*))=\varphi_n(x^*). \label{g7}
\ee
The values $f_n$ and $\varphi_n$ for different $n$ and $s$ are presented in 
Table\ref{tab1}. 
\begin{table}[p]
\caption{The dependence of the fixed point coordinates and the eigenvalues
of the transition matrix on the $s$ parameter of dividing the phase 
space into the layers and the $n$ component number of the model.}
\label{tab1}
\begin{center}
\begin{tabular}{lllllll}
\hline
~~$s$~~&$n$~&~$x^*$~~&~$E_1$~~&$E_2$~&~~$f_n$~~&~~
$\varphi_n$\\
\hline
1.1 & 1 & 10.9487 & 1.2077 & 0.9092 & 0.0181 & 0.0201 \\
    & 2 & 11.9534 & 1.2074 & 0.9089 & 0.0202 & 0.0168 \\
    & 3 & 12.8801 & 1.2072 & 0.9086 & 0.0217 & 0.0144 \\
\hline
1.5 & 1 & 3.3645 & 2.1521 & 0.6753 & 0.0998 & 0.1088 \\
    & 2 & 3.5927 & 2.1379 & 0.6710 & 0.1134 & 0.0914 \\
    & 3 & 3.8088 & 2.1273 & 0.6676 & 0.1234 & 0.0787 \\
\hline
2.0 & 1 & 1.5562 & 3.4761 & 0.5347 & 0.2153 & 0.2497 \\
    & 2 & 1.5671 & 3.3901 & 0.5260 & 0.2492 & 0.2105 \\
    & 3 & 1.5848 & 3.3256 & 0.5185 & 0.2746 & 0.1814 \\
\hline
3.0 & 1 & 0.3425 & 6.3985 & 0.4140 & 0.4640 & 0.6263 \\
    & 2 & 0.1684 & 5.9509 & 0.4010 & 0.5498 & 0.5287 \\
    & 3 & 0.0154 & 5.6298 & 0.3880 & 0.6145 & 0.4538 \\
\hline
4.0 & 1 & -0.1789 & 9.6225 & 0.3560 & 0.7167 & 1.0885 \\
    & 2 & -0.4575 & 8.5533 & 0.3402 & 0.8620 & 0.9180 \\
    & 3 & -0.7086 & 7.8304 & 0.3233 & 0.9712 & 0.7841 \\
\hline
\end{tabular}
\end{center}
\end{table}

The presence of the fixed point in RR (\ref{g4}) allows to 
write them in the linearized form 
\be
\lp
\begin{array}{cc} r_{l+1}^{(n)}-r_n^* \\
u_{l+1}^{(n)}-u_n^*
\end{array}\rp=
\Re \lp
\begin{array}{cc} r_l^{(n)}-r_n^* \\
u_l^{(n)}-u_n^*
\end{array}\rp.
\label{g11}
\ee
In calculating the matrix elements of $\Re$ matrix we restrict our 
consideration to linear on $(x_l-x^*)$ approximation. As a result the 
following expressions are obtained for the matrix elements \cite{ref44} 
\footnote 
{It should be pointed out that in (\ref{g12}) we introduce the following 
definition 
$$
 \mu_1=\mu_0\lp
a_1-\frac{q_1}{2}\rp;~~\mu_0=\frac{s^{d/2}\sqrt{n+2}U_n(y^*)}
  {\sqrt 3 \varphi_n^{1/2}(x^*)},$$
$$ \omega_0=\frac{s^{2d}\varphi_n(y^*)}{\varphi_n(x^*)};~~
  \omega_1=\omega_0(b_1-q_1),$$
$$ a_1=\tilde P_1 y^* r_1,~~r_1=\partial_1-\frac{q_1}{2},~~
 \tilde P_m=\frac{1}{U_n(y^*)}\lp\frac{d^m U_n(y_l)}
  {dy_l^{m}}\rp_{y^*}, $$
$$ b_1=\tilde Q_1 y^* r_1,~~
 \tilde Q_m=\frac{1}{\varphi_n(y^*)}\lp\frac{d^m\varphi_n(y_l)}{dy_l^{m}}
  \rp_{y^*},$$
$$ R_{12}^{(0)}=R_{12}(u_n^*)^{1/2},~~R_{21}^{(0)}=R_{21}(u_n^*)^{-1/2},\\
$$}
\bea
R_{11}=s^2\sqrt{3}\mu_1,~~R_{12}=\frac{s^2}{2}
  (\mu_0-\mu_1 x^*)(u_n^*)^{-1/2},\non
R_{21}=s^{4-d}\sqrt{3u_n^*}\omega_1,~~R_{22}=s^{4-d}\lp \omega_0-
  \frac{\omega_1 x^*}{2}\rp.\label{g12}
\eea
The eigenvalues $E_1$ and $E_2$ of the matrix $\Re$ are universal values 
\cite{ref44},\cite{ref48},\cite{ref49} 
\be
E_{1,2}=\frac{1}{2}\lb R_{11}+R_{22}\pm\lbr\lp R_{11}-R_{22}\rp^2+
  4R_{12}^{(0)}R_{21}^{(0)}\rbr ^{1/2}\rb.\label{g14}
\ee
As is obvious from Table\ref{tab1}, we have a saddle-type fixed point 
$E_1>1, E_2<1$ for all values $n$ and $s$. According to 
\cite{ref48},\cite{ref49} the bigger eigenvalue $E_1$ defines the 
critical exponent $\nu$ for the correlation length 
\be
\nu=\frac{ln~s}{ln~E_1}.\label{g15}
\ee

Since $E_1$ is a universal value, the critical exponents are also universal 
and do not depend on the microscopic characteristics of the system. As we 
can see from (\ref{g12}) and (\ref{g14}), they only depend on the 
dimensionality of space $d$ and the $n$ spin component number. The 
results of calculating the critical exponents of the model in the framework 
of this approach was shown in \cite{ref50},\cite{ref51}. Thus, in the case 
$d=3,n=3$ the following values $\nu=0.674, \alpha=-0.021, \gamma=1.347$ were 
obtained. These values of the critical exponents correspond to the $\rho^4$ 
model approximation which gives a good qualitative description of the 
critical bahaviour of the $n$-vector model. 
\footnote 
{In the marginal case of the larger $n\rightarrow \infty$, the cumulants 
$u_{2l}$ strive for their limit values	
$$\lim _{n\rightarrow \infty }u_2=1,\lim _{n\rightarrow \infty }u_4=\lim
_{n\rightarrow \infty }\frac 6{n+2}\rightarrow 0,$$
$$\lim _{n\rightarrow \infty }u_{2l}=0,\quad l=3,4,..$$
(in normalizing the spin dimensionality $m^2=n$). Considering only the 
leading by order of the value $1/n$ members in the process of integration of 
the partition functions enables us to employ the method of 
$g$-expansion \cite{ref42},\cite{ref45},\cite{ref47}. In this case the 
values $$ g_{2l}=\frac{u_{2l}}{(2l)!} / \lp \frac{d(B_1,B')}{2}\rp^l,~~~
\vec g=(g_4, g_6,...) $$
are small. It allows obtaining the next relations for the correlation length 
critical exponent (in the case $\eta=0$)
\bea
&& g^*_1=0,~~~\nu=\frac{1}{2},~~~\mbox{at}~~~ d>4;\non
&& g^*_2=\frac{1-s^{d-4}}{(n+8)(1-s^{-d})}, \non
&& \nu=\lbr 2+\frac{ln\lbr 1+\frac{n+2}{n+8}(s^{d-4}-1)\rbr}{ln~s}\rbr^{-1},
  \mbox{at}~d<4,\no
\eea
where $g^*_i$ are the fixed points. But, starting from the type of fixed 
point we have certain restrictions on the $s$ value $(1\leq s<2)$. In the 
limit $n\rightarrow\infty$ for $d<4$ we obtain $\nu=1$, which is in 
agreement with the Berlin-Kac spherical model \cite{ref43}. Analogical 
expressions for $\nu$ were obtained in a series of the works 
\cite{ref42},\cite{ref45},\cite{ref47},\cite{ref52},\cite{ref53}.}
As is known \cite{ref3}, good quantitative results for the critical 
exponents can be obtained in the framework of the $\rho^6$ model 
approximation. For example, the value of the $\nu$ critical exponent in the 
$\rho^6$ model approximation increases and practically remains unchanged 
with the increase of the $m$ order of the $\rho^{2m}$ model. Obtaining the 
eigenvectors of the transform matrix $\Re$  from (\ref{g11}) is an essential 
aspect of the investigation of the RR. They can be represented as follows 
\cite{ref1},\cite{ref44}
\be 
w_1=w_{11}\lp
\begin{array}{cc} 1\\ R_1 \end{array}\rp,~~~
w_2=w_{22}\lp
\begin{array}{cc} R\\ 1 \end{array}\rp,\label{g16}
\ee
where 
\bea
R_1=\frac{R_{21}}{E_1-R_{22}}=\frac{E_1-R_{11}}{R_{12}}, \label{g17} \\
R=\frac{R_{12}}{E_2-R_{11}}=\frac{E_2-R_{22}}{R_{21}}.\label{g18}
\eea
The conjugate vectors $v_1$ and $v_2$ are written in the form 
\be
v_1=v_{11}\lp 1~~~\frac{R_{12}}{E_1-R_{22}}\rp,~~~
v_2=v_{22}\lp \frac{E_2-R_{22}}{R_{12}}~~~1 \rp. \label{g19}
\ee
The normalization conditions $w_1v_1=1, w_2v_2=1$ give the relations for 
obtaining the coefficients $w_{ii}, v_{ii}(i=1,2)$.
\footnote 
{The relations for obtaining the coefficients $w_{ii},v_{ii}$ are written in 
the following form
$$
w_{11}v_{11}=\lbr 1+\frac{R_{12}R_{21}}{(E_1-R_{22})^2}\rbr^{-1}=
  \lbr 1+\frac{(E_1-R_{11})^2}{R_{12}R_{21}}\rbr^{-1},$$
$$ w_{22}v_{22}=\lbr 1+\frac{R_{12}R_{21}}{(E_2-R_{11})^2}\rbr^{-1}=
  \lbr 1+\frac{(E_2-R_{22})^2}{R_{12}R_{21}}\rbr^{-1}. 
$$}
Proceeding from (\ref{g11}), (\ref{g16}) the RR (\ref{g4}) can be written in 
the form 
\bea
r_l^{(n)}=r_n^*+c_1E_1^l+c_2RE_2^l, \label{g22} \\
u_l^{(n)}=u_n^*+c_1R_1E_1^l+c_2E_2^l,\no
\eea
where $c_1=c'_1w_{11};  c_2=c'_2w_{22};  c'_1, c'_2 - const$.

From the initial conditions at $l=0$
$$ r_0^{(n)}=a_2^{(n,0)}-\beta\Phi(0), u_0^{(n)}=a_4^{(n,0)},$$
the following expressions for the coefficients $á_1$ , $á_2$ were found
\bea
&& c_1=\lb r_0^{(n)}-r_n^*+(a_4^{(n,0)}-u_n^*)(-R)\rb D^{-1} \label{g23} \\
&& c_2=\lb a_4^{(n,0)}-u_n^*+(r_0^{(n)}-u_n^*)(-R_1)\rb D^{-1},\no
\eea
where 
$D=\frac{E_1-R_{11}}{R_{12}}=\frac{E_2-E_1}{E_2-R_{11}}=
\frac{E_1-E_2}{R_{11}-E_2}.$
For temperatures similar to $T_c$ the coefficients $c_1$ and $c_2$ can be 
represented as follows
\bea
&& c_1\lp T \rp=c_{1T}\tau\beta\Phi\lp 0 \rp), \label{gg24} \\
&& c_2\lp T \rp=c_{2T}\lbr \beta\Phi\lp 0 \rp \rbr^{2}. \no
\eea
Having applied these relations we evaluate the critical region of the 
temperatures $\tau<\tau^{*}$ in which the solutions of RR (\ref{g22}) are 
valid. For the critical region to exist it is necessary that the "exit" from 
this region at $l\to 1$ should not exceed the "entrance". It means that the 
value $\tau^{*}$ is equal to the smaller root of the two equations 
$$ c_{2}RE_{2}=c_{1}E_{1},\qquad c_{1}R_{1}E_{1}=c_{2}E_{2}.$$
Take into account (\ref{gg24}) we obtain for $\tau^{*}_{1,2}$
$$
\tau^{*}_{1}=|\frac{c_{2T} R E_{2}\beta "(0)}{c_{1T}E_{1}}|,\quad 
\tau^{*}_{2}=|\frac{c_{2T} E_{2}\beta "(0)}{c_{1T}E_{1}R_{1}}|.
$$
The results of calculating $\tau^{*}_{1,2}$ in the case $s=4$ are shown 
in Table\ref{tab3.1}. 
\begin{table}[htb]
\caption{The dependence of the temperature critical region $\tau^{*}_{1,2}$ 
on the $n$ spin component number of the model. }
\label{tab3.1}
\begin{center}
\begin{tabular}{lll}
$n$~&~$\tau^{*}_{1}$~&~$\tau^{*}_{2}$\\
\hline
 1 & 0.0286 & 0.3449 \\
 2 & 0.0371 & 0.3448 \\
 3 & 0.0424 & 0.3270 \\
\end{tabular}
\end{center}
\end{table}
As $c_1(T)\sim \tau$, the value $c_{1T}$ can be  represented in the 
following approximation
\be
c_{1T}=c_{1k}+c_{1k1}\tau+O\lp\tau^{2}\rp,\label{gg25}
\ee
where 
\bea
&& c_{1k}=\lbr
c_{11}+\frac{c_{12}}{\lbr\beta_c\Phi(0)\rbr^{2}}\rbr D^{-1},\label{gg26} \\
&& c_{1k1}=\frac{c_{12}}{\lbr\beta_c\Phi(0)\rbr^{2}} D^{-1},\no
\eea
with 
\bea
&& c_{11}=1-f_n-R^{*}\varphi_n^{1/2},\label{gg27} \\
&& c_{12}=-a_4^{\lp n,0 \rp}R^{*}\varphi_n^{-1/2}.\no
\eea
In accordance with (\ref{gg24}) for $c_{2T}$ we consider only the terms 
proportional to $\tau^{2}$
\be
c_{2T}=c_{2k}+\tau c_{2k1}+\tau^{2}c_{2k2}+O\lp\tau^{3}\rp,\label{gg28} \\
\ee
where the following definitions are introduced 
\footnote
{It should be pointed out that
$$
c_{21}=a_4^{(n,0)},
c_{22}=-a_2^{(n,0)} R_1^{*} \varphi_n^{1/2},$$
$$
c_{23}=R_1^{*}\varphi_n^{1/2}\lp 1-f_n \rp-\varphi_n, $$
$$
R_1^{*}=R_1\lp u_n^{*}\rp^{-1/2}, R^{*}=R\lp u_n^{*}\rp^{1/2}.$$
}
\bea
&& c_{2k}=\lbr c_{23}+\frac{c_{22}}{\beta_c\Phi(0)}+\frac{c_{21}}
{\lbr\beta_c\Phi(0)\rbr^{2}}\rbr D^{-1},\non
&& c_{2k1}=\lbr\frac{c_{22}}{\beta_c\Phi(0)}+\frac{2c_{21}}
{\lbr\beta_c\Phi(0)\rbr^{2}}\rbr D^{-1},\label{gg29} \\
&& c_{2k2}=\frac{c_{21}}{\lbr\beta_c\Phi(0)\rbr^{2}} D^{-1}.\no
\eea
The calculation of the partial partition functions (\ref{f45}),(\ref{f46})
are connected with the employment of common RR (\ref{f48}). In the vicinity 
of the fixed point (\ref{g5}) they can be replaced by the approximate 
relations (\ref{g22}) which are exact for $T=T_c$. The question arises under 
what conditions the relations (\ref{g22}) can be made use of instead of 
common RR (\ref{f48}). It is obvious that the system of the relations 
\bea
&& \vert r_l^{(n)}-r_n^{*}\vert\le\vert r_n^{*}\vert,\label{gg30} \\
&& \vert u_l^{(n)}-u_n^{*}\vert\le u_n^{*}\no
\eea
is a condition of the applicability of the approximate relations 
(\ref{g22}). 

The magnitudes which form (\ref{gg30}) are connected with the value of the 
$n$ spin component number and the value of the interaction number $l$.
The main reason for the deviation of the values $r_l^{(n)}$ and $u_l^{(n)}$ 
from their fixed values is the availability of the terms proportional to 
$c_1E_1^{l}$ in the solutions (\ref{g22}). For small values $l$ the 
contribution of these terms is small us compared with $r_n^{*}$ and 
$u_n^{*}$, as $c_1\sim\tau$. But in the case $T\not=T_c$ there always exists 
such a value as $l=m_{\tau}$ so that the contribution will be of the order 
$r_n^{*}$ or $u_n^{*}$. With $l>m_{\tau}$ the deviation will become 
considerable and the equations (\ref{g22}) cannot be used for the 
description of common RR (\ref{f48}). It is necessary to use the first 
equation in (\ref{g22}) for determining the value $m_{\tau}$. It is 
connected with the presence of the $R_1$ coefficient near the term $c_1E_1$ 
in the second equation of (\ref{g22}). Proceeding from the assumption that 
$R_1$ is small value as was shown in the case $n=1$ \cite{ref3}, we see that 
the deviation $r_l^{(n)}$ from $r_n^{*}$ will be going on faster than the 
deviation $u_l^{(n)}$ from $u_n^{*}$. Hence it follows that the value 
$m_{\tau}$ may be determined from the equation 
\be
r_{m_{\tau}+1}-r_n^{*}=\delta r_n^{*},\label{gg31} \\
\ee
where $\delta$ is a certain constant value $(\delta\le 1)$. Since later we 
will compare the results obtained in the case of $T>T_c$ with those obtained 
for $T<T_c$, we assume that $\delta=1$ with $T<T_c$ and $\delta=-1$ 
with $T>T_c$. The condition similar to (\ref{gg31}) was employed in 
\cite{ref3}.In the first approximation on $(E_2/E_1)^{\mt}$ for $\mt$ we 
have \be
\mt=-\frac{ln \mid \tau\mid}{ln E_1}+m_{0}-1,
\ee
where 
$$
m_0=m_c+m_{1}\tau.
$$
For the coefficients $m_c$ and $m_1$ the following relation take place 
$$
m_c=\frac{ln(f_n\delta/c_{1k})}{ln E_1}, m_1=-\frac{c_{1k1}}{c_{1k}ln 
E_1},$$
The obtained value of $\mt$ defines the point of exit of the system from the 
critical region 
$$
B_{\mt}=B's^{-\mt}.
$$

 The analysis of the RR shows, that in the vicinity of the PT point two
different fluctuation processes take place. The first one 
corresponds to the index values $l\in \lp 0,m_{\tau}\rp$ and describes the 
renormalization group symmetry that takes place in the vicinity of the PT 
point. It is so called critical regime (CR). The second one corresponds to 
the index values $l>m_{\tau}$ and describes the long-wavelength 
fluctuations (LWF) of the spin and is valid both near and far from the PT 
point. It is characterized by Gaussian distribution with dispersion 
depending on the availability of the CR. It is so called Gaussian regime 
for $T>T_c$ and the inverse Gaussian regime for $T<T_c$. 

The calculation of the PT temperature is a significant moment of the 
investigation of the critical behaviour of the model. As was shown in the 
works \cite{ref1},\cite{ref3}, the critical temperature is the point where 
Gaussian regime is absent and subsequent relations take place
\bea
&& \lim_{l \rightarrow \infty} r_{l+1}^{(n)}(T_c)=
\lim_{l \rightarrow \infty} r_l^{(n)}(T_c)=r_n^*=const, \label{g24} \\
&& \lim_{l \rightarrow \infty} u_{l+1}^{(n)}(T_c)=
\lim_{l \rightarrow \infty} u_l^{(n)}(T_c)=u_n^*=const.\no
\eea
In accodance with (\ref{g22}), this condition is realized only in the case 
\be
c_1(T_c)=0. \label{g25}
\ee
With (\ref{g23}) taken ino account the explicit equation for critical 
temperature was obtained 
\bea
&&\lbr \beta_c\Phi(0)\rbr^{2}\lp
1-f_n-\varphi_n^{1/2}R^{*}\rp-a_2^{(n,0)}\beta_c\Phi(0)= \non
&&-a_4^{(n,0)}R^{*}\varphi_n^{-1/2},\label{g26}
\eea
where $R^{*}=R\sqrt{u_n^{*}}$, and $a_2^{(n,0)}$, $a_4^{(n,0)}$ are 
functions of the initial interaction potential (\ref{f9}) \cite{ref54}. In 
this paper we present the rezults of calculating the PT temperature with the 
interaction potential of the (\ref{f9}) type. As follows from (\ref{f9}), 
the obtaining of the concrete calculation results are connected with the 
choice of the value $\bar{\Phi}$, that is the value $\Phi(k)$ with 
$k\in\Delta$. The correction, considering the influence of the Fourier 
transform of the potential in the interval $k\in [B^{'},B)$, makes the 
results of the calculation more precise. Let us choose the next form for 
$\bar{\Phi}$
\be
\bar{\Phi}=\langle {\Phi}(k) \rangle+{\Phi}_{\infty},\label{gn27}
\ee
where 
\be
\langle {\Phi}(k) \rangle=\frac{\int_{B^{\prime}}^{B}
{dk\Phi(k)k^2}}{\int_{B^{\prime}}^{B}
{dk k^2}}.
\ee
The results obtained for the temperature in the limit $\frac{b}{c}\to 
\infty$ must be in agreement with the mean field theory results, i.e.
$$\beta_c\Phi(0)=\frac{n}{m^2}.$$
Take into account this condition and the equation for the temperature 
$T_c$ (\ref{g26}), we obtain the equation for defining the ${\Phi}_{\infty}$.
The solution of this equation is written as 
\be
{\Phi}_{\infty}=-\frac{(n+2)(f_n+\varphi^{\frac{1}{2}}R^{*})}
{3n(1-s_{0}^{-d})}.\label{gn30}
\ee

The results of calculating the PT temperature of the $n$-vector model in the 
case $m^2=n$ are represented in Fig.1. 
\begin{figure}[thbp] 
\begin{center} 
\epsfysize=5cm 
\frame{\epsffile{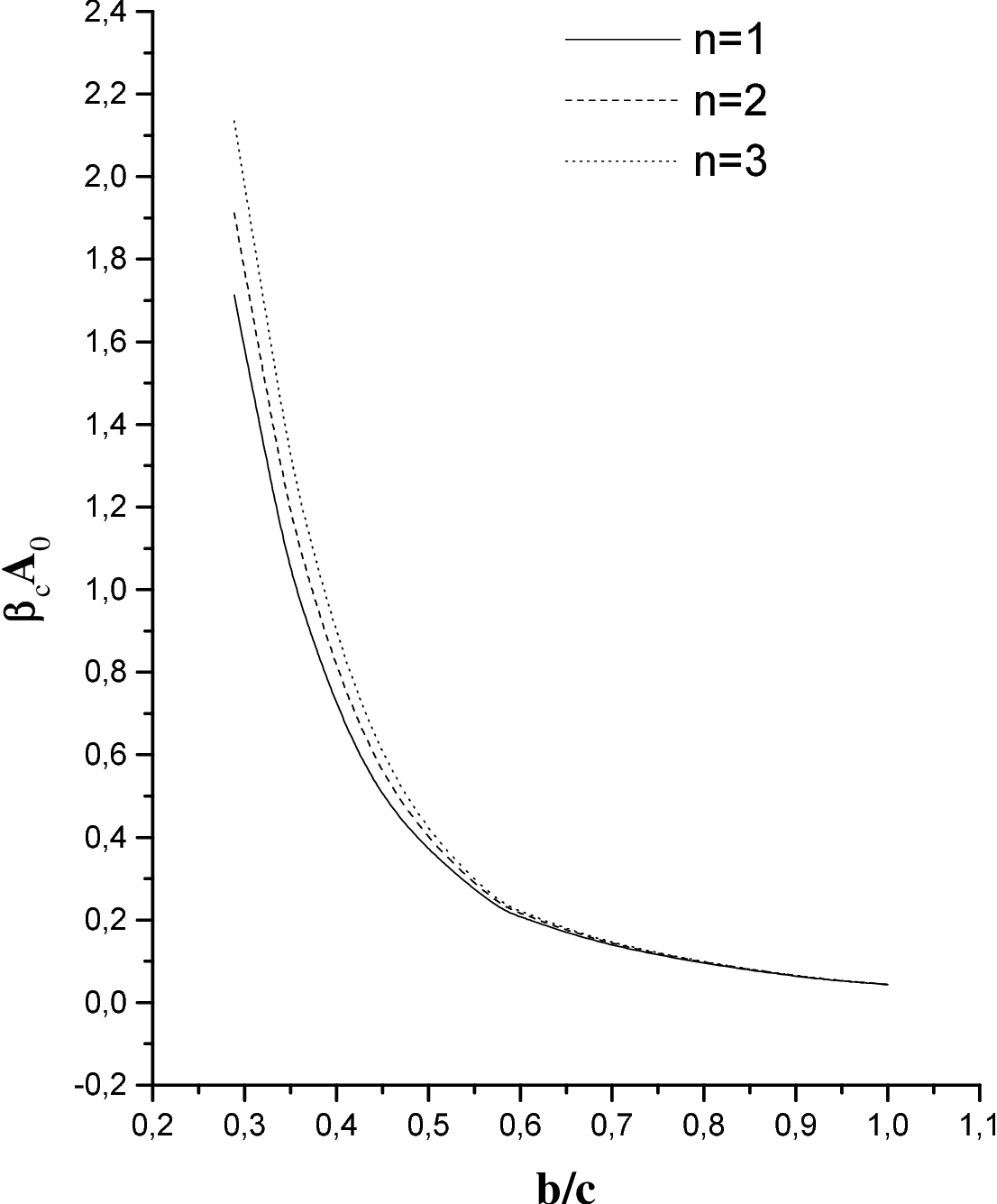}}
\end{center}
\caption{The dependence of the PT temperature on the ratio of the effective 
range $b$ of the interaction potential to the lattice constant $c$ for 
different values of the $n$ component number of the model.}
\label{à¨á6}
\end{figure}
As we can see from the figure above, the PT temperature in the 
$\beta_c A_{0}$ units (where $A_{0}$ is a constant (see (\ref{f2})) grows 
with the increase of the $n$ component number of the model. The PT 
temperature decreases and strives for the mean field theory results 
in the case of the increase of the effective range of the interaction 
potential $b$. Let us employ the introduced definitions.

\renewcommand{\theequation}{\arabic{section}.\arabic{equation}}
\section{Calculation of the free energy of the $n$-vector magnetic 
model.}
\setcounter{equation}{0}

As shown in the previous section, the calculation of the free energy of the 
system can be performed in accodance with (\ref{f51}). The given analysis of 
the RR and the definition of the region of the availability of their 
approximate solutions enable us to present the free energy of the system in 
the vicinity of the PT point in the form 
\be
F = F_{0} + F_{CR} + F_{LWF},\label{t1}
\ee
where $F_0$ is the free energy of the non-interacting spins 
\be
F_0= -kTN ln \lbr \frac{(2\pi)^{n/2}m^{n-1}}{\Gamma (n/2)}\rbr,\label{t2}
\ee
$F_{LWF}$ is the LWF free energy,
$F_{CR}$ is the CR free energy
\be
F_{CR}= -kT \sum_{l=0}^{\mu_r} F_l, \label{t3}
\ee
where 
\bea
&& F_l= N_l f_l,\non
&& f_l=\frac{n}{4} ln \lp \frac{n+2}{\varphi_n(y_l-1)} \rp+
   ln U \lp \frac{n-1}{2}, x_l\rp + \non
&& + ln U \lp \frac{n-1}{2}, y_{l-1}\rp +
\frac{x_l^2}{4}+\frac{y^2_{l-1}}{4},\label{t4}
\eea
Here $N_l=N's^{-dl}$. In the case $l=0$ for $f_l$ the following relation 
takes place
\bea
&& f_0=\frac{n}{4} ln \lp \frac{3}{u'_4} \rp+u'_0+\frac{3}{4} \frac{(u'_2)^2}
  {u'_4}+ ln U \lp \frac{n-1}{2}, z'\rp + \non
&& + \frac{n}{4} ln \lp \frac{3}{u_0^{(n)}}\rp +
\frac{x_0^2}{4}+ln U \lp \frac{n-1}{2},x_0\rp.\label{t5}
\eea
The employment of the RR solutions (\ref{g22}) allows us to select in $f_l$ 
the explicit dependence on the number of the phase layer $l$. Having 
performed the summation along the layers of the CV phase space to the point 
$m_{\tau}$ of the exit of the system from the CR according to (\ref{t3}), 
for the free energy of the CR we obtain 
\be
F_{CR}^{\pm}=-kTN'\lbr 
\gamma'_{01}+\gamma_1\tau+\gamma_2\tau^2-
   \gamma_{10}^{\pm}\mid\tau\mid^{d\nu}\rbr.\label{t6}
\ee
Let note that the signs $"+"$ and $"-"$ correspond to $T>T_c$ and $T<T_c$ 
respectively. The coefficients $\gamma'_{01},\gamma_1,\gamma_2, 
\gamma_{10}^{\pm}$ are constants and do not depend on the temperature (see 
Appendix 3). The analytical part of the CR free energy is connected with the 
coefficients $\gamma'_{01},\gamma_1,\gamma_2$. It should be mentioned that 
the expressions of these coefficients coincide at the temperatures above and 
below the critical temperature. The dependence of $\gamma_l$ on the 
microscopic parameters of the interaction potential and the $n$ component 
number of the model are shown in Table\ref{tab2.2}.
\begin{table}[p]
\caption{The dependence of the coefficients of the free energy on the 
microscopic parameters of the interaction potential and the 
characteristics of the crystal lattice for different $n$ component number of 
the model.}
\label{tab2.2}
\vspace{0.5cm}
\begin{center}
\begin{tabular}{llllllllll}
~$\!\!\!b/c\!\!\!$~&~$n\!\!\!$~&~$\gamma_{01}^{'}$~&~$\gamma_1$~&~$\gamma_2$~&~$\gamma_{10}^+$~&~
$\gamma_{10}^-$~&~$\gamma_0\!\!\!$~&~$\gamma_3^+$~&~$\gamma_3^-$\\
\hline
$\!\!\!0.2887\!\!\!$ & $1\!\!\!$ & $0.349\!\!$ &-0.500 &-0.459 & -0.538 &  
2.737 
 & $1.811\!\!\!$ & 1.283& 2.726\\
    &$ 2\!\!\!\!$ & $0.727\!\!$ &-0.976 &-4.427 & 3.315 &  0.368  & 
    $5.335\!\!\!$ & 5.795 &6.294\\
    & $3\!\!\!$ & $1.099\!\!$ &-1.435 & 24.497 &-25.359 & 30.049  & 
    $9.488\!\!\!$&-22.603 & -22.530\\
\hline
$\!\!\!b=c\!\!\!$ & $1\!\!\!$ & $0.297\!\!$ &-0.521 &-0.122 &-0.448 &  
2.276 & 
 $61.085\!\!\!$ & 1.066 & 2.266\\
    & $2\!\!\!$ & $0.620\!\!$ &-1.011 &-3.200 & 2.722 &  0.303 & 
    $192.194\!\!\!$ & 4.759 &5.170\\
    & $3\!\!\!$ & $0.938\!\!$ &-1.470 & 21.176 &-21.397 & 25.358 & 
    $349.648\!\!\!$&-19.072 & -19.013\\
\end{tabular}
\end{center}
\end{table}
The nonanalytical part, characterizing the temperature dependence of the 
specific heat in the vicinity of the PT point, is connected with the term 
$\gamma^{\pm}_{10}\tau^{d\nu}$, where 
\be
\gamma^{\pm}_{10}=\bar{\gamma}' s^{-d\mu_0},\label{t7}
\ee
\be
\bar{\gamma}'=\frac{f^*_{CR}}{1-s^{-d}}-\frac{f_n\delta d_1}{1-s^{-d}E_1}+
   \frac{f_n^2\delta^2 d_3}{1-s^{-d}E_1^2}.\label{t8}
\ee
For $f^*_{CR}$ we have   
\be
f^*_{CR}=\frac{n}{2}\ln 
y^{*}+\tilde{\alpha}(y^{*})^{-2}+\frac{(x^{*})^2}{4}+\ln 
U(\frac{n-1}{2},x^{*}).\label{t9}
\ee
The values of $d_m$ are given in Appendix 4. In Table 2 the dependence of 
$\gamma_{10}^{\pm}$ on the component number $n$ of the model for different 
ratios $b/c$ is shown. The expression (\ref{t6}) describes the contribution 
of the region of the renormalization group symmetry to the free energy of 
the system. It allows us to obtain the respective contributions to the 
specific heat of the system at $T<T_c$ and $T>T_c$
\be
C^{\pm}_{CR}=kN'\lbr 
c^{(0)}-c_{CR}^{\pm}\mid\tau\mid^{-\alpha}\rbr,\label{t10}
\ee
where 
\bea
&& \alpha=2-d\nu,\non
&& c^{(0)}=2(\gamma_1+\gamma_2),\\ \label{t11}
&& c_{CR}^{\pm}=(1-\alpha)u^{\pm}_{CR},\non
&& u^{\pm}_{CR}=d\nu\gamma_{10}^{\pm}.\no
\eea
The curves 1 in Fig.2 correspond to the contribution of the CR to the 
specific heat at $T<T_c$ and $T>T_c$ respectively. The negative value of the 
specific heat amplitude, that corresponds to the contribution of the CR, 
testifies to the nonstability of the contribution of the short-wavelength 
fluctuations (SWF) of the spin moment density. Considering the contribution 
of the region of the LWF of the spin moment density in calculating the 
thermodynamic characteristics of the system is a topical problem of to-day.
\begin{figure}[thbp] 
\begin{center} 
\epsfysize=5cm 
\frame{\epsffile{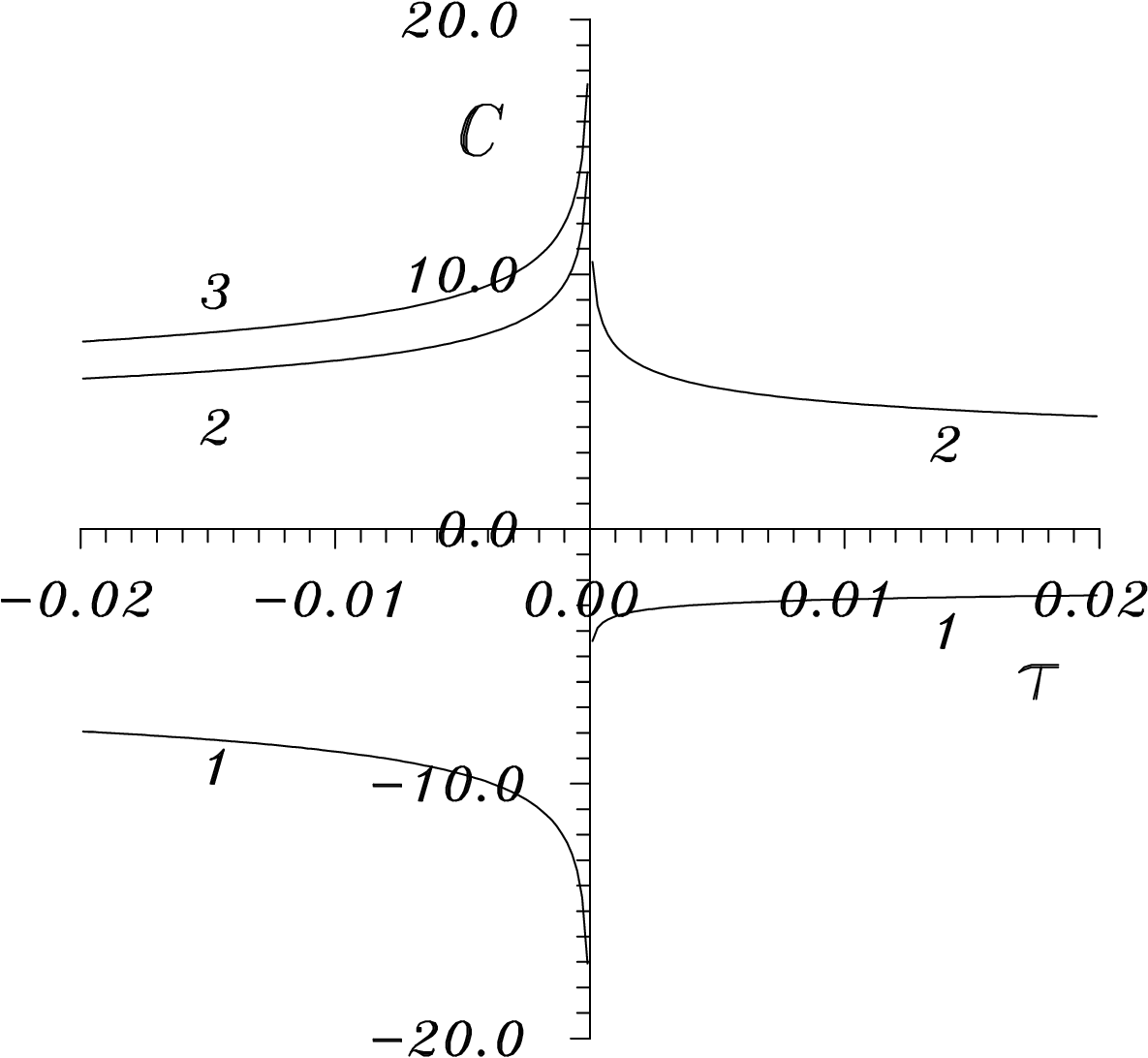}}
\end{center}
\caption{The temperature dependence of the specific heat. The comparison of 
the contribution of the CR and the region of the LWF. Notes: 1 - corresponds 
to the CR, 2 - limit Gaussian region (LGR) ($\tau>0$) and inverse Gaussian 
region (IGR) ($\tau<0$), 3 - contribution to the specific heat due to the 
rise of the ordering in the system.} 
\label{à¨á1}
\end{figure}

The regions of LGR for $T>T_c$ and IGR for $T<T_c$ correspond to the LWF 
of the spin moment density. The increase of the basic $x_{l}$ and the 
intermediate $y_{l}$ variables as functions of $l$ is a typical 
peculiarity of the LGR and IGR. In this connection the contribution of 
these regions to the free energy can be calculated in the Gaussian measure 
density approximation (as in this region of the wave vectors the 
term proportional to the fourth power under the exponent of the 
distribution functions becomes much smaller than the square term). Since, 
the increase of $x_{l}$ is gradual, there exists the so-called transition 
region (TR) where it is necessary to keep the fourth power of the 
$\vrho_{\bf{k}}$ variables in the distribution function. The value of the 
TR is defined by a certain number of the CV phase space layers $m^{''}$ 
following the point $m_{\tau}$ of the exit of the system from the CR. The 
value of the TR is defined by the condition  \be
\mid x_{\mt+m_{0}^{''}}\mid = \frac {\alpha_{m}}{1-s^{-d}},\label{t12}
\ee
where $\alpha_{m}$ is a constant ($\alpha_{m}\geq 10$). The arguments 
mentioned above anable us to write the contribution of the LWF at $T>T_c$ 
to the free energy of the system in the form 
\be
F_{LGR}^{+}=-kTN^{'} f^{+}_{LGR}\tau^{3\nu} - \beta\mu_{B}^{2} 
H^{2}N\gamma_{4}\tau^{-2\nu}, \label{t13}
\ee
where  
$$ 
f^{+}_{LGR}=f_{TR}+f^{'}_{*}.\label{t14}
$$
It should be pointed out that the following definitions are introduced here:
$\mu_{B}$ is Bor's magneton, $f_{TR}$ corresponds to the contribution of the 
TR and $f^{'}_{*}$ corresponds to the contribution of the wave vectors 
region with $k<B^{'}s^{-(\mt+m^{''}+1)}$. The explicit analytical 
expressions of these coefficients was shown in \cite{ref50}. For $\gamma_{4}$
we have  
\be
\gamma_{4}=s^{2m^{''}}(\frac{f_{n}\delta}{c_{1T}})^{2\nu}\frac{1}{2\beta 
\Phi(0)g_{0}},\label{t15}
\ee
where $g_{0}$ is a complex function of $x^{*}_{m^{'}_{\tau}-1}$, with  
$\mt^{'}=\mt+m^{''}+2$ (see.\cite{ref50}).

The contribution of the region of the LWF of the spin moment density at  
$T<T_c$ to the free energy of the system is described by the value 
$F^{-}_{LGR}$. The value $F^{-}_{LGR}$ corresponds to the inverse Gaussian 
region. The calculation of the inverse Gaussian region contribution has its 
own peculiarities. In the temperature region $T<T_c$ the large-scale 
fluctuations of the spin moment density are described by the non-Gaussian 
distribution in which the coefficient near the square term becomes negative. 
This indicates the appearance of the nonzero order parameter of the system. 
After selecting the ordering free energy the distribution of the 
fluctuations becomes Gaussian (see Appendix 5). As a result for 
$F^{-}_{LGR}$ we obtain 
\be 
F^{-}_{LGR}=-kTN' 
|\tau|^{3\nu}\gamma^-_{LGR},\label{t16}\ee
where 
\bea
&& \gamma_{LGR}^-=\gamma_3^{(\mut)}+\gamma_3^{<\vec{\sigma}>},\\
&& \gamma_3^{(\mut)}=\gamma_g+\gamma_{\rho},\quad 
\gamma_3^{<\vec{\sigma}>}=c^d_{\nu}\bar{\gamma}_3^{<\vec{\sigma}>},\quad
\bar{\gamma}_3^{<\vec{\sigma}>}=s_0^d E'_0.\no
\eea
The value $\gamma_3^{<\vec{\sigma}>}$ corresponds to the contribution from 
the ordering in the system. The coefficients $\gamma_g$ and $\gamma_{\rho}$ 
are written in the form 
\bea
&& \gamma_g=\bar{\gamma}_g c^d_{\nu},\\
&& \bar{\gamma}_g=ln \lbr \lp\frac{s^{d+4}(n+2)\bar 
u_{\mut}}{3\pi^2\varphi_n(x_{\mut})}
   \rp^{n/4} e^{\frac{y^2_{\mut}}{4}}U \lp \frac{n-1}{2}, y_{\mut}\rp\rbr\no
\eea
\bea
&& \gamma_{\rho}=\bar{\gamma}_{\rho} c^d_{\nu},\\
&& \bar{\gamma}_{\rho}=\frac{5}{12}n L(x) -\frac{n}{2} ln \lp \frac{1+2\bar 
r_{\mutb}}{\pi}\rp +\frac{n}{3} -\non
&& -\frac{\bar u_{\mutb}}{8} \lp 
\bar{\alpha}_1^2n^2+3n^3L(x)\bar{\alpha}_2\rp+\non
&& +\frac{\bar u_{\mutb}^2}{48}n^4 \lp \bar{\alpha}_4+3\bar{\alpha}_1^2
   \bar{\alpha}_2\rp +\frac{9}{4} n^2\bar{\alpha}_2\bar r_{\mutb}^2.\no
\eea
It should be pointed out that the following definitions are introduced here 
\be
\bar r_{\mt+1}=f_n(1+\delta),\quad \bar u_{\mt+1}=\varphi_n-f_n 
\varphi_n^{1/2} R_1^* \delta.
\ee
\be
L(x)=3\lp\frac{x-arctg x}{x^3}\rp,
x = \frac{1}{\sqrt {2 \bar r_{\mt+1}}},
\ee
besides, $c_{\nu}=(\frac{c_{1T}}{f_{n}\delta})^{\nu}$ is a nonuniversal 
value connected with the microscopic parameters of the initial Hamiltonian.
According to (\ref{t1}), (\ref{t6}), (\ref{t13}) and (\ref{t16}) the 
complete expression of the free energy of the three dimensional 
$n$ - vector magnetic model in the absence of the external field can be 
written as 
\be
F=-kTN'\lbr \gamma_0-\gamma_1|\tau|+\gamma_2|\tau|^2
   +\gamma_3^{\pm}|\tau|^{3\nu}\rbr,\label{t17}
\ee
where 
\bea
\gamma_0=\gamma'_{01}+s_0^3 ln \lbr \frac{(2\pi)^{n/2}m^{n-1}}{\Gamma(n/2)} 
\rbr,\\
\gamma_3^{\pm}=\gamma^{\pm}_{LGR}-\gamma_{10}^{\pm}.
\eea

The coefficient $\gamma_3^{\pm}$ includes the contribution of the CR and the 
region of the LWF of the spin moment density at the temperatures above and 
below the critical. It describes the singular behaviour of the specific heat 
in the vicinity of the PT point. The dependence of the coefficients 
$\gamma_0, \gamma_3^{\pm}$ on the microscopic parameters of the interaction 
potential and the characteristics of the crystal lattice for different $n$ 
component number of the model is shown in Table\ref{tab2.2}.

\renewcommand{\theequation}{\arabic{section}.\arabic{equation}}
\section{Thermodynamic functions of the $n$-vector magnetic 
model in the framework of the $\rho^{4}$ model approximation.}
\setcounter{equation}{0}

The obtained complete expression for the free energy of the $n$-vector 
magnetic model (\ref{t17}) in the $\rho^4$ model approximation allows us to 
calculate other thermodynamic functions in the vicinity of the phase 
transition point. As was remarked above, (see (\ref{t10}), Fig.2), in 
calculating the thermodynamic functions of the system especially 
significant is considering the contribution of the LWF of the spin moment 
density. So, differentiating the expression of the free energy by the 
temperature, we obtain the expression for the entropy 
\be 
S=kN\lbr S^{0}+S^{1}\tau+u_3^{\pm}|\tau|^{3\nu-1}\rbr,\label{ff1}
\ee 
where 
$$
S^{0}=\gamma_0+\gamma_1, S^{1}=2(\gamma_1+\gamma_2),
u_3^{\pm}=\pm 3\nu\gamma_3^{\pm}.\label{ff2}
$$
The values of the coefficients $S^{0}, S^{1}, u_3^{\pm}$ for different $n$ 
values are shown in Table\ref{tab2.3}.
\begin{table}[htbp]
\caption{The dependence of the amplitudes of the thermodynamic functions on 
the microscopic parameters of the interaction potential and the 
characteristics of the crystal lattice for different $n$ component number 
of the model.}
\label{tab2.3}
\vspace{0.5cm}
\begin{center}
\begin{tabular}{lllllllll}
$b/c$~&~$n$~&~$S^{0}$~&~$S^{1}$~&~$u^{+}_3$~&~$u^{-}_3$~&~$C^{(0)}$~&~$C^{+}_1$~&~$C^{-}_1$\\
\hline
0.2887 & 1 & 1.311 & -1.917 & 2.355 & -5.007 & -1.917 & 1.972 & 4.190\\
       & 2 & 4.359 & -10.805 & 11.226 & -12.196 & -10.805 & 10.529 & 11.436\\
       & 3 & 8.053 & 46.124 & -45.677 & 45.530 & 46.124 & -46.629 & -46.478\\
\hline
b=c & 1 & 60.564 & -1.286 & 1.958 & -4.163 & -1.286 & 1.639 & 3.484\\
    & 2 & 191.183 & -8.421 & 9.220 & -10.017 & -8.421 & 8.646 & 9.393\\
    & 3 & 348.178 & 39.412 & -38.540 & 38.422 & 39.412 & -39.344 & -39.223\\
\end{tabular}
\end{center}
\end{table}

It should be mentioned that the correct temperature behaviour of the 
specific heat curves is ensured by considering the region of the LWF of the 
spin moment density, i.e. the region of LGR at $T>T_c$ and the region of 
IGR at $T<T_c$. The significant characteristic of the system is the 
specific heat for which we obtain 
\be
C=kN'\lbr C^{(0)}+C_1^{\pm}|\tau|^{-\alpha}\rbr,\label{ff3}
\ee 
where 
$$
\alpha=2-3\nu,\\
C^{(0)}=2(\gamma_1+\gamma_2), C_1^{\pm}=3\nu(1-\alpha)\gamma_3^{\pm}.
$$
The second term in (\ref{ff3}) describes the main peculiarity of the 
specific heat behaviour in the vicinity of the PT point. As we can see from 
(\ref{ff3}), the coefficient $C_1^{\pm}$ includes the contributions of the 
CR and region of the LWF of the spin moment density. We can see from 
Fig.2, that considering the influence of the regions of LWF (curves 2) 
provides the positivity of the specific heat and the system stability 
respectively. The dependence of the coefficients $C^{(0)}$ and $C_1^{\pm}$ 
on the microscopic parameters of the Hamiltonian, i.e. from the $b/c$ ratio 
for different $n$ is exhibited in Table\ref{tab2.3}.

In Fig.3 the temperature dependence of the specific heat for different $n$ 
is shown. As it was noticed above, in the case $n=3$ the critical exponent 
$\alpha$, which describes the singularity of the specific heat becomes 
negative $\alpha=-0.021$ \cite{ref50},\cite{ref51}. The analysis of the 
expression (\ref{ff3}) and the obtained values of the specific heat 
amplitudes (see Table\ref{tab2.3}) shows that the specific heat in the 
case $n=3$ does not diverge and receives a concrete finite value (see 
Fig.3). 
\begin{figure}[thbp] 
\begin{center} 
\epsfysize=5cm 
\frame{\epsffile{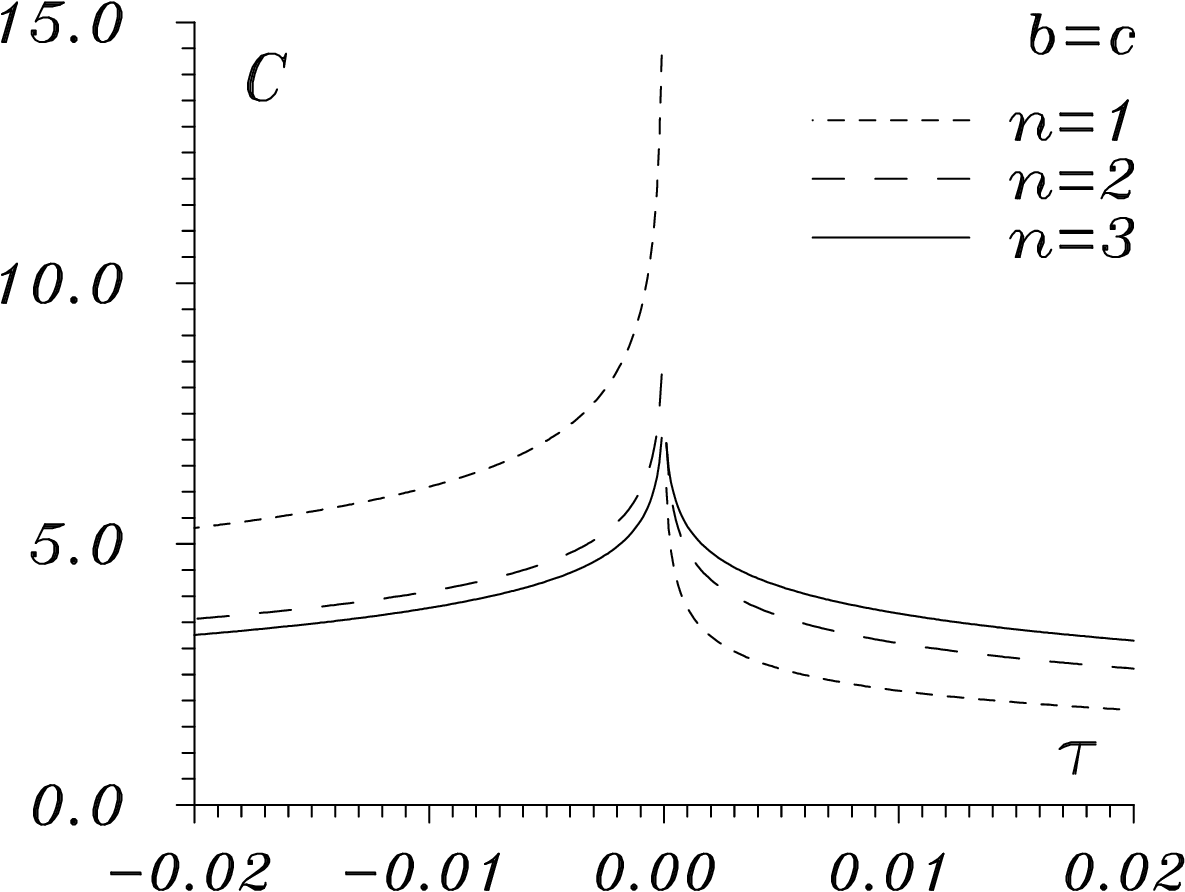}}
\end{center}
\caption{The temperature dependence of the specific heat for different $n$ 
component number of the model.}
\label{à¨á7}
\end{figure}

\begin{figure}[thbp] 
\begin{center} 
\epsfysize=5cm 
\frame{\epsffile{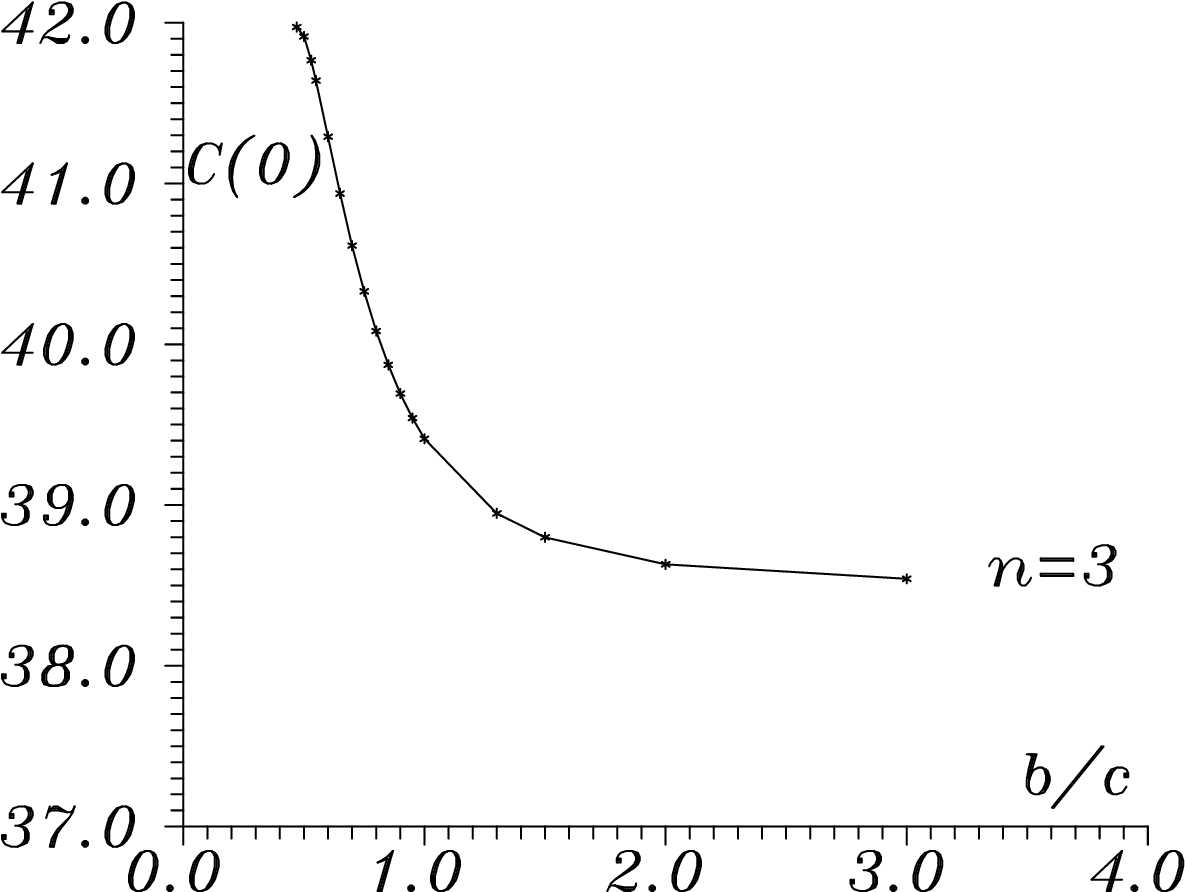}}
\end{center}
\caption{The dependence of the maximum of the specific heat on the ratio of 
the effective range $b$ of the interaction potential to the crystal lattice 
constant $c$ in the case $n=3$ and $T=T_c$.}
\label{à¨á8}
\end{figure}

The curve of the dependence of the specific heat maximum at $T=T_c$ in the 
case $n=3$ on the ratio of the effective range $b$ of the interaction 
potential to the lattice constant $c$ is shown in Fig.4. One can see from 
this figure that the value of the specific heat maximum decreases and 
tends to cnstant value as the effective range $b$ of the interaction 
potential increases. This agrees with the results of the mean field theory. 

According to (\ref{f3}) the ratio of the critical amplitudes of the 
leading singular terms of the specific heat at $T>T_c$ and $T<T_c$ can be 
written in the form 
\be
A=\frac{\gamma_3^+}{\gamma_3^-}.
\ee

The comparison of the results obtained for the ratio of the critical 
amplitudes of the specific heat leading singular terms with results 
obtained by other methods is shown in Table\ref{tab2}.
\begin{table}[hbt]
\caption{The dependence of the ratio of the specific heat critical 
amplitudes at $T>T_c$ and $T<T_c$ on the $n$ component number of the 
model.} 
\label{tab2}
\vspace{0.5cm}
\begin{center}
\begin{tabular}{lllllll}
$ n $~&~$A (This $~&~$\epsilon-exp $~&~$Pade-$~&~$\frac 1s-exp 
$~&~$experi-$\\
&~$ study) $~&~$ \cite{ref15} $~&~$ apr \cite{ref22} $~&~$ \cite{ref46} 
$~&~$ ment\cite{ref55} $\\
\hline
 1 & 0.470 & 0.55 & 0.438 & 0.519 & 0.538 \\
 2 & 0.921 & 0.99 & 0.880 & 0.888 & 1.067 \\
 3 & 1.003 & 1.36 & 1.326 & 1.152 & 1.588 \\
\end{tabular}
\end{center}
\end{table}
\section*{Conclusions}

In general, it should be noted that the separate accounting of the 
contributions of the short- and long- wavelength fluctuations of the spin 
moment density in the expression of the free energy of the system in the 
vicinity of the PT temperature allows us to find the explicit analytical 
expressions for the thermodynamic functions as functions of the 
temperature. The proposed method enables as to investigate the dependence 
of the critical amplitudes of the thermodynamic functions on the 
microscopic parameters of the interaction potential and the characteristics 
of the crystal lattice. The results obtained for the critical exponents and 
the ratio of the critical amplitudes agree with those obtained by other 
methods. The negligible deviation of the obtained results from the 
experimental data and the results of the numerical calculations is 
connected with the restriction in the calculation by the $\rho^4$ model 
approximation. As was seen in \cite{ref3},\cite{ref56}, the employment of 
the $\rho^6$ measure density for the investigation of the PT in the case 
$n=1$ gives a more precise definition of the calculation results of the 
universal and nonuniversal characteristics of the system. The 
extension of the suggested method to the investigation of the critical 
behaviour of the $n$-vector model in the frame of the $\rho^6$ model 
approximation does not require any principal changes. To be short, in the 
present paper we do not give enough attention to calculating the scaling 
corrections. In the case $n=1$ such calculations were made in \cite{ref3}. 
Besides, in the common case the critical exponent of the correlation 
function is $\eta \neq 0$. Such values of the correlation function 
critical exponent can be obtained by the method we propose if we take into 
account the correction on the averaging of the interaction potential in 
each layer of the CV. We are going to perform such investigations in our 
subsequent papers.

\renewcommand{\theequation}{A1.\arabic{equation}}
\section*{Appendix 1}
\setcounter{equation}{0}

The $\vrho_{\bf {k}}$ CV are introduced by means of the functional 
representation for the operators of the spin density fluctuation 
$\hat{\vrho}$
$${\hat{\vrho}}_{\QTR{bf}{k}}=\int{\vrho
 }_{\QTR{bf}{k}}J\left(\vrho -\hat{\vrho} \right) \left(
 d{\vrho}_{\QTR{bf}{k}}\right) ^N, 
 {\hat{\vrho}_{\QTR{bf}{k}}}=\frac 
 1{\sqrt{N}}\sum_{\QTR{bf}{R}}\QTR{bf}{\hat S}_{\QTR{bf}{R}}\exp\left( 
 -i\QTR{bf}{kR}\right) , $$
where  
$$J\left( \bf{\vrho -\hat {\vrho} }\right) =\left[
 \prod_{\bf{k}}^{\prime }\delta 
 \left(\vrho_{\bf{k}}^c-\hat{\vrho}_{\bf{k}}^c\right) 
 \delta \left(\vrho_{\bf{k}}^s-\hat{\vrho} 
 _{\bf{k}}^s\right)\right] \delta \left( \vrho_0-\hat{\vrho}_0\right) 
 - $$ 
 is the transition operator. For the CV $\vrho_{\bf{k}}$ the following 
 relations come into being 
$$\vrho_{\QTR{bf}{k}}=\vrho
_{\QTR{bf}{k}}^c-i\vrho_{\QTR{bf}{k}}^s,\quad \vrho_{\bf {k}}^{c} = 
\vrho_{-\bf {k}}^{c},\quad \vrho_{\bf {k}}^{s} = -
\vrho_{-\bf {k}}^{s}.$$ 
 
The Jacobian of the transition from spins to CV has the form 
\cite{ref3},\cite{ref44} 
\be
J[\rho] =\tilde Z_0\int \exp (2i\pi\sum_{\bf{k}} 
\vrho_{\bf{k}}\vomeg_{\bf{k}}+\bar D [\omega]) 
 (d\vomeg_{\bf{k}})^{N}, \label{f4}
\ee
where for $\vomeg_{\bf{k}}$ variables conjuagate to $\vrho_{\bf{k}}$ 
variables we have 
$$\vomeg_{\QTR{bf}{k}}=\frac 12\left(\vomeg
_{\QTR{bf}{k}}^c+i\vomeg_{\QTR{bf}{k}}^s\right), \quad \vomeg_{\bf 
{k}}^{c}=\vomeg_{-\bf {k}}^{c},\quad \vomeg_{\bf {k}}^{s}=-\vomeg_{-\bf 
{k}}^{s}$$
 $$(d\vomeg_{\bf{k}})^N=\prod\limits_{a=1}^nd\omega 
 _0^a\prod\limits_{\bf{k}}^{^{\prime }}d\omega 
 _{\bf{k}}^{a,c}d\omega_{\bf{k}}^{a,s}, $$
and 
 \EQN{\tilde Z_0=\left[ \frac{\left( 2\pi \right) ^{\frac n2}m^{n-1}}{\Gamma
 \left( \frac n2\right) }\right] ^N.}{f5}

For the values $\bar D [\omega] $ the following  relation is valid
 \EQN{\bar D [\omega] =\sum_{l\geq
 1}\sum_{\QTR{bf}{k}_1,...\QTR{bf}{k}_{2l}}\frac{\left( 2\pi i\right)
 ^{2l}}{\left( 2l\right) !}\frac{u_{2l}}{N^{l-1}}\vomeg
 _{\bf{k}_1}...\vomeg_{\bf{k}_{2l}}\delta
 _{\QTR{bf}{k}_1+...+\QTR{bf}{k}_{2l}}\quad .}{f6}

It should be noted that here and further by the product of the vectors of 
the $\vomeg_{\bf{k}_{1}}...\vomeg_{\bf{k}_{2l}}$ type we understand the sum 
of various possible scalar products of given vectors. The coefficients 
$u_{2l}$ have the following expressions \cite{ref42} 
\be
 u_2=\frac{m^2}n,\qquad u_4=-\frac{6m^4}{n^2(n+2)},\label{f7}
\ee
$$ u_6=15m^6 [\frac 1{n (n+2)(n+4) }-\frac
 3{n^2(n+2) }+\frac 2{n^3}].$$

\renewcommand{\theequation}{A2.\arabic{equation}}
\section*{Appendix 2}
\setcounter{equation}{0}

For the coefficients $a_{2l}$ the following expressions are available 
 \be
 a_2=\frac n{m^2}s_0^{\frac d2}\left[ \frac 1{1-\frac{20m^2}{\left(
 n+4\right) }\sum\limits_{k\in \Delta}\frac{\beta \Phi (k) }N}\right] 
 ^{\frac 12}U_0, \label{f22}
 \ee
 $$ 
 a_4=-\frac{3n^2}{m^4}s_0^d [ \frac 1{1-\frac{20m^2}{(n+4) 
 }\sum\limits_{k\in \Delta }\frac{\beta \Phi (k)}N}]
 ( 1-z^{^{\prime }}U_n (z^{^{\prime}})-U_0^2) .
 $$
  Here subsequent definitions are introduced 
 \bea 
 U_0=\sqrt{\frac{n+2}2}U_n\left( z^{^{\prime }}\right),\qquad
 \Delta =[B^{^{\prime }},B),\label{f23}
 \eea
 \EQN{U_n\left( z^{^{\prime }}\right) =\frac{U\left(
 \frac{n+1}2,z^{^{\prime }}\right) }{U\left( \frac{n-1}2,z^{^{\prime
 }}\right) },\qquad z^{^{\prime }}=\sqrt{\frac 3{u_4^{^{\prime
 }}}}u_2^{^{\prime }},}{f24}

where $U\left( a,x\right) =D_{-a-\frac 12}\left( x\right) $ are the Weber 
cylinder parabolic functions. The renormalized $u_{2l}$ cumulants 
considering the availability of $\Phi(k)$ for the large values of $k$ have 
the form 
  $$u_0^{^{\prime }}=s_0^d\frac{u_2n}{2N}\sum_{k\in \Delta }\beta
  \Phi (k) ,\quad u_2^{^{\prime }}=u_2-\frac{\left|
  u_4\right| n}{2N}\sum_{k\in \Delta }\beta \Phi (k) , $$
  \EQN{u_4^{\prime }=\left[ \left| u_4\right|
  -\frac{u_6n}{2N}\sum_{k\in \Delta }\beta \Phi (k) \right] s_0^{-d}.}{f15}

\renewcommand{\theequation}{A3.\arabic{equation}}
\section*{Appendix 3}
\setcounter{equation}{0}
For the coefficients $d_i$ the following expressions are valid
\bea
d_1&=&B_3\lp A_3+\frac{A_1}{E_1}\rp,~~~
d_2=B_1\lp A_3+\frac{A_1}{E_2}\rp,~~~
d_3=B_6\lp A_3+\frac{A_1}{E_1^2}\rp+B_3^2\lp A_4+\frac{A_2}{E_1^2}\rp\non
d_4&=&B_2\lp A_3+\frac{A_1}{E_2^2}\rp+B_1^2\lp A_4+\frac{A_2}{E_2^2}\rp,~~~
d_5=B_4\lp A_3+\frac{A_1}{E_1E_2}\rp+2B_1B_3\lp 
A_4+\frac{A_2}{E_1E_2}\rp\non
d_6&=&B_7\lp A_3+\frac{A_1}{E_1E_2^2}\rp+2(B_1B_6+B_3B_4)\lp A_4+
\frac{A_2}{E_1^2 E_2}\rp,\non
d_7&=&B_5\lp A_3+\frac{A_1}{E_1E_2^2}\rp+2(B_1B_4+B_2B_3)\lp A_4+
\frac{A_2}{E_1 E_2^2}\rp,\non
d_8&=&B_8\lp A_3+\frac{A_1}{E_1^2E_2^2}\rp+2\lp B_1B_7+B_2B_6+B_3B_5+
\frac{B_4^2}{2}\rp\lp A_4+A_2E_1^{-2}E_2^{-2}\rp,\nonumber
\eea
where the coefficients $A_{i},B_{j}$ are universal constants and are 
defined by the fixed point coordinates. For the $B_{j}$ coefficients we 
have 
\bea
B_1&=&\varphi_n^{-1}\lp R^* \sqrt 3-\frac{x^*}{2}\rp,~~~
B_2=\varphi_n^{-2}\lp \frac{3}{8}x^*-\frac{\sqrt 3}{2}R^*\rp,~~~
B_3=\varphi_n^{-1/2}\lp \sqrt 3-\frac{R_1^*x^*}{2}\rp\non
B_4&=&\varphi_n^{-3/2} \lp -\frac{\sqrt 3}{2} R_1^*R^*-\frac{\sqrt 3}{2}+
\frac{3}{4} R_1^* x^*\rp,~~~
B_5=\varphi_n^{-5/2} \frac{3\sqrt 3}{4}
\lp R_1^*R^*+\frac{1}{2}-\frac{5}{4\sqrt 3}R_1^* x^*
\rp,\non
B_6&=&\varphi_n^{-1} R_1^* \lp \frac{3}{8} x^* R_1^*
-\frac{\sqrt 3}{2}\rp,~~~
B_7=\varphi_n^{-2}\frac{3\sqrt 3}{4} R_1^* \lp 1+ \frac{R_1^* R^*}{2}
-\frac{5R_1^* x^*}{4\sqrt 3}\rp,\non
B_8&=&\frac{\varphi_n^{-3}15\sqrt 3 R_1^*}{16}\lp \frac{7x^* R_1^*}{4\sqrt 3}
-(1+R^*R_1^*)\rp.\nonumber
\eea
The coefficients $A_{i}$ are written in the form 
$$
A_1=\frac{n}{2} r_1-\frac{2r_1\tilde\alpha}{(y^*)^2},~~~
A_2=\frac{n}{2} r_2-\frac{n}{4}r_1^2+\frac{\tilde\alpha(3r_1^2-2r_2)}
{(y^*)^2},~~~
A_3=-\frac{n}{2} U_n (x^*),~~~
A_4=-\frac{n}{4} U'(x^*),\nonumber
$$
where the following definition are introduced
\bea
\tilde\alpha&=&\frac{(5n+16)n}{8}-\frac{n/2(n/2+1)}{2},~~~
U'_n(x^*)=\lp\frac{dU_n(x_l)}{dx_l}\rp_{x^*},\non
r_1&=&\tilde d_1-\frac{q_1}{2},~~~r_2=\frac{1}{2}\tilde d_2-\frac{1}{2} 
\tilde d_1 q_1+\frac{3}{8} q_1^2-\frac{1}{4} q_2,\non
\tilde d_i&=&\frac{1}{U_n(x^*)}\lp\frac{d^i U_n(x_l)}{dx_l^i}\rp_{x^*},~~~
q_i=\frac{1}{\varphi_n(x^*)}\lp\frac{d^i\varphi_n(x_l)}{dx_l^i}
\rp_{x^*}\nonumber
\eea

\renewcommand{\theequation}{A4.\arabic{equation}}
\section*{Appendix 4}
\setcounter{equation}{0}
The $\gamma_{0i}$ coefficients are expressed in the form 
\bea
\gamma_{01}&=&s^{-d}\lp \frac{f^*_{Š}}{1-s^{-d}}+\frac{c_{2k}d_2E_2}
{1-s^{-d}E_2}+\frac{c^2_{2k}d_4 E_2^2}{1-s^{-d}E_2^2}\rp,\non
\gamma_{02}&=&s^{-d}\lp \frac{c_{2k1}d_2E_2}{1-s^{-d}E_2}+
\frac{b_1d_4E_2^2}{1-s^{-d}E_2^2}+
\frac{c_{1k}d_1E_1}{1-s^{-d}E_1}+
\frac{c_{1k}c_{2k}d_5E_1E_2}{1-s^{-d}E_1E_2}+
\frac{c_{1k}b_0d_7E_2^2E_1}{1-s^{-d}E_1E_2^2}\rp,\non
\gamma_{03}&=&s^{-d}\lp \frac{c_{2k2}d_2E_2}{1-s^{-d}E_2}+
\frac{(c_{1k}c_{2k1}+c_{2k}c_{1k1})d_5E_1E_2}{1-s^{-d}E_1E_2}+
\frac{c_{1k1}d_1E_1}{1-s^{-d}E_1}+
\frac{b_2d_4E_2^2}{1-s^{-d}E_2^2}+\rd\non
&+&\ld\frac{(c_{1k}b_1+c_{1k1}b_0)d_7E_1E_2^2}{1-s^{-d}E_1E_2^2}+
\frac{c_{1k}^2 d_3E_1^2}{1-s^{-d}E_1^2}+
\frac{c_{1k}^2c_{2k}d_6E_1^2E_2}{1-s^{-d}E_1^2E_2}+
\frac{c_{1k}^2 b_0d_8E_1^2E_2^2}{1-s^{-d}E_1^2E_2^2}\rp,\nonumber
\eea
where 
$$
b_0=c^2_{2k},~~~b_1=2c_{2k}c_{2k1},~~~b_2=c^2_{2k1}+2c_{2k}c_{2k2}.\nonumber
$$
For $\gamma^{'}$ we have 
$$
\gamma'=\frac{f^*_{Š}}{1-s^{-d}}+\frac{f_n\delta d_1}{1-s^{-d}E_1}+
\frac{f_n^2\delta^2 d_3}{1-s^{-d}E_1^2},\nonumber
$$
where a definition is introduced 
$$
f_n=\tau E_1^{m_\tau+1} c_{1k}.\nonumber
$$

\renewcommand{\theequation}{A5.\arabic{equation}}
\section*{Appendix 5}
\setcounter{equation}{0}
The contribution of the IGR into the free energy can be written in the form 
\be
F^{-}_{LGR}=-kTN' s^{-d(\mt+1)} ln \lp 2^{n/2}Q(P_{\mt})\rp - kT ln 
Z_{\mt+1},\label{a5.1}
\ee
where 
\bea
&& Z_{\mt+1}=\int exp \lbr -\fr \sum_{k\leq B_{\mt+1}} d^{(n,\mt+1)}(k)
   \rhok\rhomk -\rd \non
&& \ld \frac{a_4^{(n,\mt+1)}}{4!N_{\mt+1}}
   \sum_{{\bf{k_1,...k_4}}\atop{k_i\leq
   B_{\mt+1}}}\rhokf\delf\rbr(d\vrho)^{N_{\mt+1}},\non
&& Q(P_{\mt})=(2\pi)^{-n/2}\lb \frac{n+2}{3} s^{d} \frac{a_4^{(n,\mt-1)}}
{\varphi_{n}(x_{\mt-1})}\rb ^{n/4}\times \non 
&& U(\frac{n-1}{2},y_{\mt-1})\exp({\frac{y_{\mt-1}^2}{4}}). \label{a5.2}
\eea
After the separation of the ordering free energy in $Z_{\mt+1}$ the 
coefficient at the square term becomes positive. It allows us to use under 
the integration of $Z_{\mt+1}$ by the variables $\vrho_{k}$ with 
${\bf{k}}\ne 0$ the Gaussian measure density as basic. After the 
integration for $Z_{\mt+1}$ we obtain 
\bea
&& Z_{\mt+1}=e^{-\beta F_{\mt+1}} \int exp \lp \beta \sqrt N \mu_B \vec H 
\vrho_0+ \right. \non
&& +\left.B\vrho_0^2-\frac{G}{N} \vrho_0^4 \rb d\vrho_0.\label{a5.3}
\eea
The value $F_{\mt+1}$ corresponds to the contribution of the region of the 
LWF of the spin moment density (i.e. from the variables $\vrho_{\bf{k}}$ 
with $k\to 0$, but $k\neq 0$) into the free energy of the system
\bea
&& -\beta F_{\mt+1}=N_{\mt+1} \lb \frac{3}{2} \mid d^{(n,\mt+1)}(0)\mid
   I_1-\rd \non
&& - \frac{n}{2}\frac{1}{N_{\mt+1}}\sum'_{k\leq B_{\mt+1}}
   ln \lp\frac{\frac{}{d_2(k)}}{\pi}\rp - \non
&& -\frac{a_4^{(n,\mt+1)}I_1^2}{8}+\frac{a_4^{2(n,\mt+1)}}{48}
   (I_4+3I_1^2I_2)+\non
&& +\ld \frac{9}{4} \mid d^{(n,\mutb)}(0)\mid^2 I_2-
   \frac{b^{2(n,\mutb)}}{8} I_1I_2\rb.\label{a5.33}
\eea
For $I_{i}(i=1,...,4)$ the next relations take place 
\bea
&& I_1=n\bar{\alpha}_1\frac{s^{2(\mutb)}}{\beta\Phi(0)},\quad \bar{\alpha}_1=
   \frac{L(x)}{2\bar{r}_{\mutb}},\non
&& I_2=n^2\bar{\alpha}_2 \lbr \frac{s^{2(\mutb)}}{\beta\Phi(0)}\rbr^2,\quad 
   \bar{\alpha}_2=(\bar{\alpha}_1^2+6e_1^2(1+e_2^2)),\\
&& I_3=n^3\bar{\alpha}_3 \lbr \frac{s^{2(\mutb)}}{\beta\Phi(0)}\rbr^3,\quad 
   \bar{\alpha}_3=\bar{\alpha}_1^3+6e_1^3\lp 1+\frac{e_2^3}{\sqrt 2}\rp,\non 
&& I_4=n^4\bar{\alpha}_4 \lbr \frac{s^{2(\mutb)}}{\beta\Phi(0)}\rbr^4,\quad 
   \bar{\alpha}_4=\bar{\alpha}_1^4+6e_1^4\lp 1+\frac{e_2^4}{\sqrt 2}\rp.\no
\eea
It should be mentioned, that 
\bea
&& e_1=\frac{15}{\pi^2 [3+10\bar r_{\mutb}]},\\
&& e_2=\frac{1}{2\pi}\lb sin (\pi\sqrt 2)-\pi\sqrt 2 cos (\pi\sqrt 2)\rb.\no
\eea
The variable $\vrho_0$ in (\ref{a5.3}) is connected with order parameter. 
Its mean value is proportional to the spin density of the system. For the 
values $B$ and $G$  we obtain 
\bea
&& B=\mid \tau \mid^{2\nu}B_0, B_0=\beta\Phi(0)B^{(0)},\\
&& B^{(0)}=c^2_{\nu} f_n \frac{1+\delta}{2} B_1^{(0)},\non
&& B_1^{(0)}=1-\alpha_{11}\bar{u}_{\mutb}+\alpha_{22}
   \frac{\bar{u}_{\mutb}^2}{2\bar{r}_{\mutb}},\no
\eea
\bea
&& G=\mid \tau \mid^{\nu}G_0, G_0=[\beta\Phi(0)]^2 G^{(0)},\\
&& G^{(0)}=c_{\nu} \frac{s_0^3}{24} \bar{u}_{\mutb}g^{(0)},\quad 
g^{(0)}=1-\frac{3}{2}n^2 \bar{\alpha}_{2}\bar{u}_{\mutb},\no
\eea
where 
\be
\alpha_{11}=\frac{n}{2} \frac{\bar{\alpha}_1}{\bar r_{\mutb}}+\frac{3}{2} n^2
   \bar{\alpha}_2, \quad
\alpha_{22}=n^3 \lb \frac{\bar{\alpha}_1\bar{\alpha}_2}{2}+
   \frac{\bar{\alpha}_3}{3}\rb.
\ee
The variable $\vrho_0$ is a macroscopic value, so we can accept that 
\be
\vrho_0=\sqrt N \vrho.\label{a5.4}
\ee
It makes it possible for us to apply the saddle - point method for the 
calculation of $Z_{\mt+1}$ in (\ref{a5.3}). As a result we find  
\be
Z_{\mt+1}=\sqrt {\frac{2\pi}{E''_0(<\vrho>)}} exp \lb -\beta F_{\mt+1}-
   N E_0 (<\vrho>)\rb,\label{a5.4b}
\ee
where $<\vrho>$ is an extreme point of expression 
\be
E_0(\vrho)=G\vrho^4-B\vrho^2-\beta \mu_ \vec H\vrho,\label{a5.5}
\ee
which arrises in (\ref{a5.3}) with the change of variables (\ref{a5.4}). 
The variable $\vrho_0$ corresponds to the operator 
$\hat{\vrho_0}=\frac{1}{\sqrt N}\sum_l\hat{\vec{\sigma_l}}$, the mean value 
of which is connected with the equilibrium value of the order parameter 
$\sigma$. In the case of $\vec H=0$ we find for $<\vrho>$ the following 
solutions 
\be
<\vrho_{1,2}>=\pm \sqrt{\frac{B}{2G}}, \quad <\vrho_{3}>=0.\label{a5.6}
\ee
The solutions $<{\vrho}_{1,2}>$ and $<\vrho_3>$ correspond to extreme value 
of the functional $E_{0}(\vrho)$ (\ref{a5.5}). The presence of the 
nonzero mean spin moment at the temperatures $T<T_c$ testifies to the 
appearance of spontaneous magnetization in the system in the absence of 
the external field. Fig.5 shows the temperature dependence of the order 
parameter $\sigma$ for different values of $n$ component number of the 
model in the absence of the external magnetic field. 
\begin{figure}[thbp] \begin{center} 
\epsfysize=4cm 
\frame{\epsffile{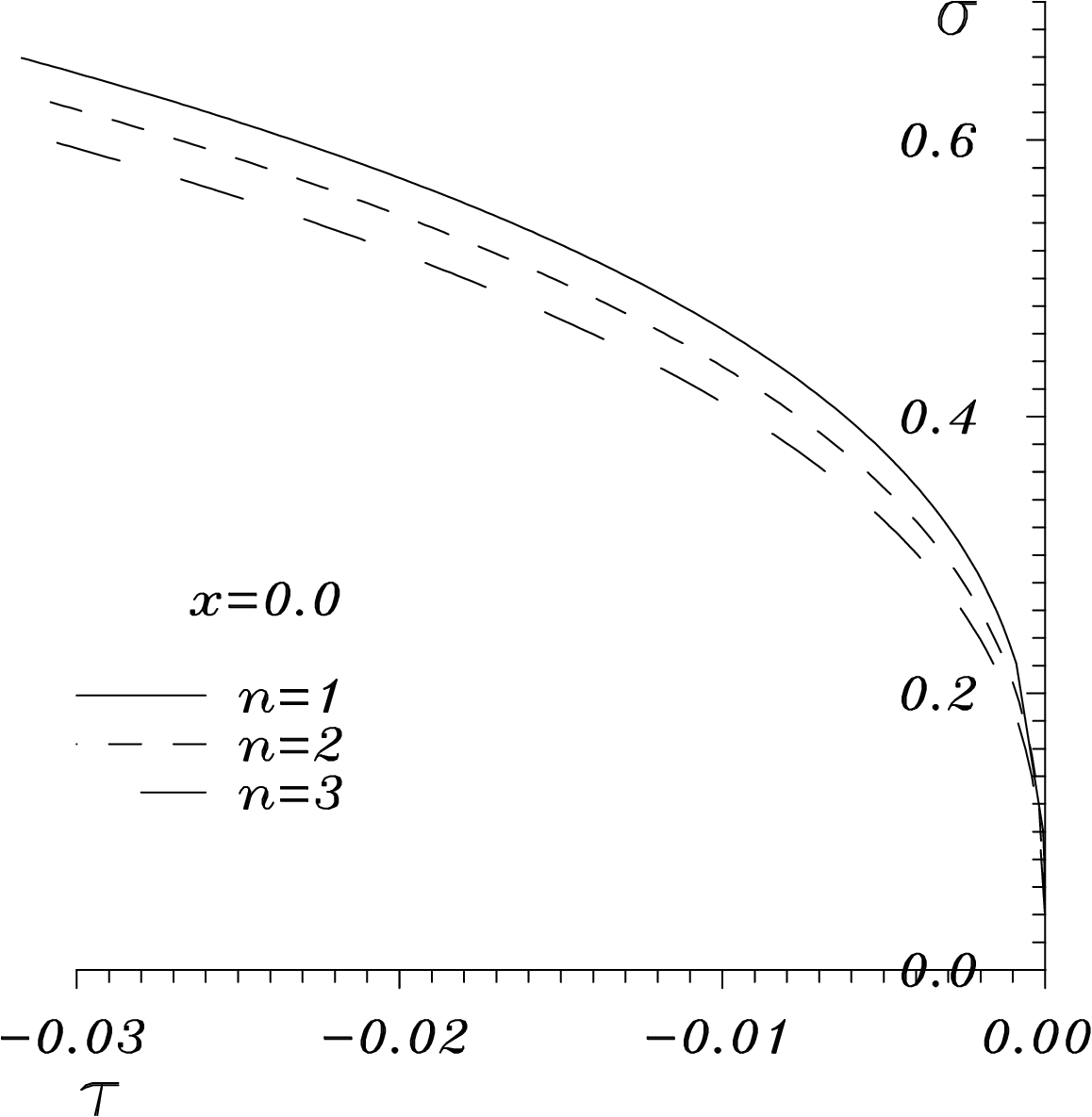}}
\end{center}
\caption{The temperature dependence of the order parameter for different 
values of the $n$ component number of the model in the case ${\vec H}=0$ 
(with $b=c$). }
\label{à¨á2}
\end{figure}

In accordance with the above mentioned for $E_0(<\vrho>)$ we obtain
\be
E_0(<\vrho>)=-\frac{B^2}{4G} = E'_0 \mid \tau \mid^{3\nu},\label{a5.7}
\ee
where 
\be
E'_0=-\frac{3}{2} \frac{\bar r_{\mutb}^2}{\bar u_{\mutb}}
   \frac{(B_1^{(0)})^2}{s_0^3 g^{(0)}}.\label{a5.8}
\ee
All this enables us to obtain the explicit analytical expression for the 
IGR free energy. Summarizing the expression (\ref{a5.33}) by the 
variables $k\leq B_{\mutb}$ $(k\neq 0)$ and considering the contribution 
of the $\vrho_0$ variables in accordance with (\ref{a5.4b}) - 
(\ref{a5.8}), for  $F^{-}_{LGR}$ we obtain the expression (\ref{t16}).

\end{document}